\documentclass[manuscript]{aastex}

\slugcomment{RCW175}

\shorttitle{RCW175}
\shortauthors{Battistelli et al.}
\usepackage{amsmath}
\usepackage{gensymb}

\begin{document}

\def\dustem{\textsc{DustEM}}
\def\Planck{\textit{Planck}}
\def\Herschel{\textit{Herschel}}
\def\Spitzer{\textit{Spitzer}}
\def\IRAS{\textit{IRAS}}

\title{New radio observations of anomalous microwave emission in the HII region RCW175}

\author{E.S. Battistelli\altaffilmark{1}, E. Carretti\altaffilmark{2}, A. Cruciani\altaffilmark{1}, P. de Bernardis\altaffilmark{1}, R. Genova-Santos\altaffilmark{3},
S. Masi\altaffilmark{1}, A. Naldi\altaffilmark{1}, R. Paladini\altaffilmark{4}, F. Piacentini\altaffilmark{1}, C.T. Tibbs\altaffilmark{4,5}, L. Verstraete\altaffilmark{6}, N. Ysard\altaffilmark{6}}

\altaffiltext{1}{Department of Physics, Sapienza University of Rome, Piazzale Aldo Moro 5, 00185 Rome, Italy}
\altaffiltext{2}{CSIRO Astronomy and Space Science, PO Box 76, Epping, NSW 1710, Australia}
\altaffiltext{3}{Instituto de Astrofisica de Canarias, C/ Vía Lactea, s/n, E38205 - La Laguna (Tenerife), Spain}
\altaffiltext{4}{Infrared Processing Analysis Center, California Institute of Technology, Pasadena, CA 91125, USA}
\altaffiltext{5}{Scientific Support Office, Directorate of Science and Robotic Exploration, European Space Research and Technology Centre (ESA/ESTEC), Keplerlaan 1, 2201 AZ Noordwijk, The Netherlands}
\altaffiltext{6}{IAS, CNRS (UMR 8617), Universit\'e Paris-Sud 11, b\^atiment 121, 91400 Orsay, France}

\begin{abstract}

We have observed the HII region RCW175 with the 64-m Parkes telescope at 8.4~GHz and 13.5~GHz in total intensity, 
and at 21.5~GHz in both total intensity and polarization. High angular resolution ranging from 1~arcmin to 
2.4~arcmin, high sensitivity, and polarization capability enable us to perform a detailed study of the different 
constituents of the HII region. For the first time, we resolve three distinct regions at microwave frequencies, two of 
which are part of the same annular diffuse structure. Our observations enable us to confirm the presence of anomalous microwave emission from RCW175. 
Fitting the integrated flux density across the entire region with the currently available spinning dust models, 
using physically motivated assumptions, indicates the presence of at least two spinning dust
components: a warm component ($T_{gas}=5800$~K) with a relatively large hydrogen number density $n_{H}$ = 26.3/cm$^{3}$ and a 
cold component ($T_{gas}=100$~K) with a hydrogen number density of $n_{H}$ = 150/cm$^{3}$. The present study is an example highlighting the potential of using high 
angular-resolution microwave data to break model parameter degeneracies. 
Thanks to the spectral coverage and angular resolution of the Parkes observations, we have been able to derive one of the first anomalous microwave emission/excess maps, at 13.5~GHz, showing clear evidence that the bulk of the anomalous emission arises in particular from one of the source components, with some additional contribution from the diffuse structure.
A cross-correlation analysis with thermal dust emission has shown a high degree of correlation with one of the regions within RCW175. In the center of RCW175, we 
find an average polarized emission at 21.5~GHz of $2.2\pm0.2 (rand.) \pm0.3 (sys.)$\% of the total emission, where we have included
both systematic and statistical uncertainties at 68\% CL. This polarized emission could be due to sub-dominant synchrotron emission 
from the region and is thus consistent with very faint or non-polarized emission associated with anomalous microwave emission.

\end{abstract}


\section{Introduction}

HII regions play a crucial role in broadening our understanding of the mechanisms regulating the interstellar medium (ISM). 
The complex morphology of HII regions is the result of an ensemble of factors, i.e. the evolutionary stage, the number and spectral 
type of the ionizing sources, and their interplay with the surrounding medium. RCW175 is a relatively well-studied, complex HII region,  
and is approximately $10\times15$ arcmin$^{2}$ in size. The region was identified by \cite{sha59} and has been observed at various 
wavelengths ever since. In the microwave range it was first observed by the VSA (Very Small Array) experiment \citep{tod10}, 
which found evidence of an anomalous microwave emission (AME) excess, and then by the CBI (Cosmic Background Imager) experiment, 
which confirmed the anomalous excess and a basic structure of at least two components at $\sim$30~GHz \citep{dic09}.

AME was first identified as diffuse emission in the 10--60~GHz range in excess of the expected level of microwave emission from free--free, 
synchrotron, and thermal dust emission by experiments such as the Cosmic Background Explorer-DMR \citep[COBE-DMR;][]{kog96}, the Saskatoon
experiment \citep{deo97}, Owens Valley Radio Observatory \citep[OVRO;][]{lei97} and the Tenerife experiment \citep{deo99}. 
Several authors have found further statistical evidence of AME 
\cite[see e.g., ][]{ban03,lag03,deo04,fin04,dav06,hil07,lu12,miv08}, while it has also been detected directly in 
individual regions in a limited number of cases \cite[see e.g.,][]{wat05,sca07,cas08,ami09,mur10,dic10,gen11,pla11a,pla14a}.
Recent models predict that the AME is dominated by electric dipole emission from the smallest grains, possibly polycyclic 
aromatic hydrocarbons \citep[PAHs;][]{dra98,ysa10,hoa10}. However, other physical emission mechanisms, such as hot
free--free emission \citep{lei97}, hard synchrotron radiation \citep{ben03}, or magnetic dipole emission \citep{dra99} cannot be completely ruled out. 
 
To date there have been only a handful of polarization observations of individual AME regions, all of which show evidence of little or no polarization: to 
the level of a few per cent \citep{rub12, bat06,mas09,dic07,cas08,hoa13, lop11,dic11}. AME polarization is able to differentiate between the different models 
since different models predict a different polarization percentage. This is the case, for instance, of magnetic dipole 
emission, which predicts a polarization fraction that can be of the order of 10\% or more with frequency-dependent behavior. In contrast, Stokes 
I multi-band observations in the 5--30~GHz range are essential to separate the AME from the other 
types of ISM emission in conjunction with mm observations for thermal dust removal. High angular resolution measurements 
are fundamental to understanding the physics behind the AME and to limit source confusion \cite[see e.g.][]{bat12}.  In this respect, though, it is remarkable
that, up until now, no observation has had sufficient sensitivity, resolution, or frequency coverage to disentangle fully the 
numerous candidate mechanisms. Arcminute-level angular resolution observations are crucial to disentangling different 
contributions within the same Galactic region. These are starting to reveal cases in which the dust-to-radio correlation 
is observed to decrease when we go to finer angular scales \cite[see e.g., ][]{vid11,cas11,tib13}. \cite{ysa11} and others have 
suggested that this behavior could be due to the fact that whereas the infrared (IR) dust emission is proportional to the 
intensity of the radiation field and to the dust column density, it is not as straightforward for spinning dust emission 
since its excitation processes are more numerous and complex. Indeed, the rotation of interstellar grains is both 
excited and damped by the emission of IR photons (proportional to the radiation field intensity), collisions with neutrals 
(HI, H$_{2}$ ) and ions (HII, CII), plasma drag (HII, CII), photoelectric emission, formation of H$_{2}$ on the grain surface, and
the emission of electric dipole radiation itself \citep{dra98}.

A detailed multi-wavelength study of RCW175 has been performed by \cite{tib12} based on publicly available data. \cite{tib12} 
computed the star formation rate of the region, identified the position of young stellar object candidates and estimated 
the total dust mass of RCW175. By combining high angular resolution \Spitzer~ \citep{wer04} and \Herschel~ \citep{pil10} IR data with the dust model \dustem~
\citep{com11}, \cite{tib12} characterised the dust properties within the region. They cross-correlated the coarse angular resolution microwave data (e.g. CBI) with the derived dust property maps, and found that the AME is not strongly correlated with the smallest dust grain abundance but rather with the interstellar radiation field within the region. A missing key  piece of information in the available data set are observations in the range 8--25~GHz, where the AME is expected to rise. These 
are essential for a detailed study of the physical characteristics of the AME, as well as for separating the various types of emission. 

In this paper we present high angular resolution, multi-frequency, and polarization-sensitive observations of this source at
8.4, 13.5, and 21.5~GHz conducted with the Parkes Radio Telescope. These observations cover the gap in the spectral energy density 
of currently available data and can increase our understanding of the morphology of RCW175 at frequencies where the AME is a 
significant fraction of the total emission. This paper is structured as follows: in section \ref{parkes} we describe the 
observations we have conducted with the Parkes telescope; in section \ref{sed} we discuss the Spectral Energy Density (SED) 
derived from our new Parkes observations in addition to other data available in literature. In section \ref{morphology} we discuss the morphology of the region, including the spatial distribution of the AME; in section \ref{ir} we present the results obtained from cross-correlating 
\Herschel~and \Spitzer~IR maps with microwave maps, while in section \ref{pol} we present the results obtained from our 
polarization observations. In section \ref{con} we discuss the interpretation of the results and present our conclusions.

\section{Observations and data reduction}\label{parkes}

The observations were conducted at three different frequency bands with the Parkes Radio Telescope, NSW Australia, 
a 64-m telescope operated as a National Facility by ATNF-CASS, a division of CSIRO. Photometric observations at 
8.4~GHz and 13.5~GHz are described in section \ref{sec:8.4obs} and section \ref{sec:13.5obs}, respectively, 
while the polarimetric observations at 21.5~GHz are described in section \ref{sec:21.5obs}.

\subsection{8.4~GHz observations}\label{sec:8.4obs}

The 8.4~GHz observations were conducted with the MARS receiver of the Parkes telescope on 2011 July 26 for 
4~hr. The receiver is a circular polarization system with $T_{\rm sys} \sim 30$~K, a resolution of 
FWHM = 2.4~arcmin and a bandwidth of 400~MHz centered at a frequency of 8.4~GHz. To detect the 
whole useful bandwidth the backend Digital Filter Banks Mark 3 (DFB3) was used with a configuration bandwidth of 
1024~MHz and 512 frequency channels each of width 2~MHz.

The correlator has full Stokes parameter capability, recording the two autocorrelation products $RR^*$, $LL^*$ 
and the complex cross-product of the two circular polarizations $RL^*$ whose Real and Imaginary parts are the two Stokes parameters Q and U. The feed illuminates the primary mirror with an edge taper of 19~dB, the first side lobe amplitude is $-30$~dB relative to the antenna pattern peak, and the gain is 1.18~Jy~K$^{-1}$.

The flux density scale was calibrated using the source PKS B1934-638 with an accuracy of 5\% \citep{rey94}. This source is compact
compared to the beam size, and in the case of slightly extended sources, as for RCW175 at this 
resolution, some flux can be picked up by the first side lobes resulting in an error of the flux scale accuracy.  
To check this we have compared the solid angle of the main beam and first sidelobes. Our calibration observations are 
not sensitive enough to measure the sidelobes, but their shape can be estimated in the case of a mirror illuminated by 
a Gaussian feed\footnote{A Gaussian is a good approximation of modern feed horn profiles.} as a function of its 
edge taper and the central blockage of the focus cabin \citep{gol87}. Following \cite{gol87}, we find that the first
sidelobe stretches out to 11~arcmin (diameter), which covers most of our source in one dimension and 
exceeds it in the other, and its solid angle is 0.6\% of the main beam. The effects on the flux accuracy are thus 
negligible. This confirms the usual assumption that it is safe to use a flux calibration performed using compact sources 
for objects extending up to a few beam-widths. The main beam efficiency is 92\%, consistent with the results 
from \cite{gol87} for similar edge taper values.

The channels spanning the IF were then binned into twenty 20-MHz sub-bands for the subsequent map-making 
processing. A standard basket-weaving technique with orthogonal scan sets along R.A. and Dec.\ spaced by 
45~arcsec was used to observe an area of $30'\times30'$ centered on the source. The scan speed was 
$0.5^\circ$/min with a sampling time of 0.25~s. Map-making software based on the \cite{eme88} Fourier 
algorithm was applied to make the maps \citep{carretti10}.  This technique effectively reduces 1/\textit{f} noise, and
removes stripes and features different in the two orthogonal sets of scans. 

\begin{figure}
\plotone{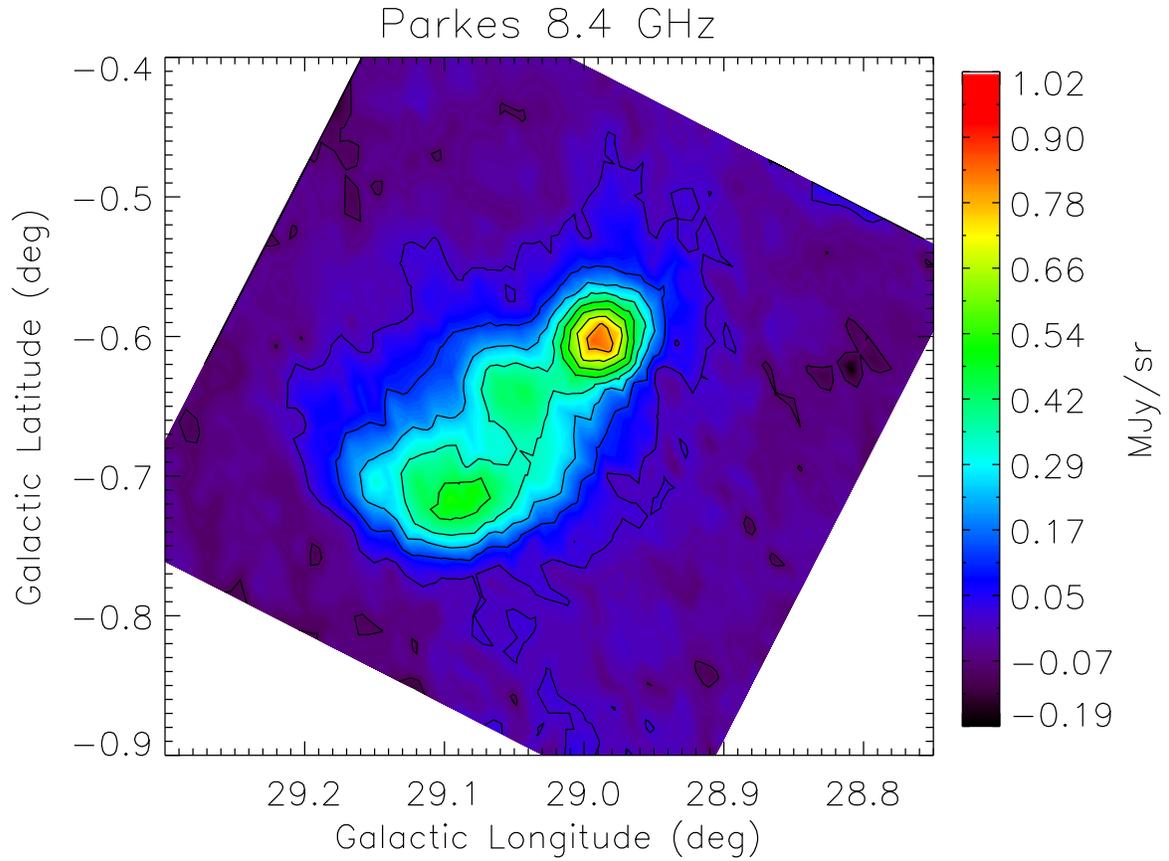}
\caption{8.4~{GHz} map of RCW175 obtained with the Parkes radio telescope using the MARS receiver. The angular 
resolution is 2.4~{arcmin}
FWHM and sensitivity is 22~{mJy/beam}.  The contour intervals are 10\% of the peak emission. \label{map}}
\end{figure}

The twenty sub-band maps were then binned together into one map for the analysis (see Figure \ref{map}). The final
 rms on the map is 22~mJy/beam on beam-size scales. This is higher than the expected sensitivity 
(some 0.7~mJy/beam). However, RCW175 is close to the Galactic plane at ($l$,~$b$) = (29.1$^\circ$, $-0.7^\circ$) and 
such an excess rms signal might be due to diffuse Galactic emission. To investigate this option we have used the 
map at 1.4~GHz and resolution of 14.5~arcmin of CHIPASS \citep{cal14}, where we find that an rms 
signal of rms$_{1.4\ \rm GHz}^{14.5'} = $700~mK in the area around RCW175. The behavior of the {rms} 
signal with the angular scale is described by {rms}~$\propto \sqrt{\ell(\ell+1)C_\ell}$, where $C_\ell$ is 
the angular power spectrum and the multipole $\ell$ relates to the angular scale $\theta$ as $\theta \sim 180^\circ/\ell$.
The angular power spectrum follows a power-law $C_\ell \propto \ell^{-1.7}$  near the Galactic plane \citep{bru02}. 
In the 1.4--8.4~GHz range the  free--free emission is the leading contribution close to the Galactic plane 
with a brightness temperature that goes as $T_b^{\rm ff} \propto \nu^{-2.1}$ as a function of the frequency $\nu$. 
Rescaling the CHIPASS rms for the frequency and angular spectrum previously described, and converting the temperature rms for the calculated gain,  we estimate a Galactic {rms} signal of {rms}$_{8.4\ \rm GHz}^{2.4'} = 
$25~{mJy/beam} at the frequency and resolution of our observations, consistent with the 
{rms} signal we detect around the source.

\subsection{13.5~{GHz} observations}\label{sec:13.5obs}

The 13.5~{GHz} observations were conducted with the Ku-band receiver of the Parkes telescope on 
2011 August 31 and 2011 September 1 for a total of 6~hr. The receiver is a linear polarization system with 
$T_{\rm sys} \sim 150$~{K}, a resolution of FWHM = 1.7~arcmin, and a bandwidth of 700~{MHz} 
centered at a frequency of 13.55~{GHz}. To detect the whole useful bandwidth the backend Digital Filter 
Banks Mark 3 (DFB3) was used with  a configuration bandwidth of 1024~{MHz}  and 512 frequency channels 
(2~{MHz} each). Only the two autocorrelation products were used ($XX^*$, $YY^*$) for Stokes~I measurements.
The feed illuminates the primary mirror with an edge taper of 18~dB, the first sidelobe amplitude is $-27$~dB 
relative to the antenna pattern peak, and the gain is 1.55{~Jy~K}$^{-1}$.

The flux density scale was calibrated using the source PKS B1934-638 with an accuracy of 5\% \citep{rey94}. As for
 the 8.4~{GHz} data set, we could not measure the secondary lobes, and to estimate the error in the flux 
scale accuracy for our extended source we estimated the antenna pattern following \cite{gol87}. In this case, to 
cover most of the source, the first two sidelobes have to be considered that stretch out to 13~{arcmin} 
(diameter). We found that their combined solid angle is 1.9\% of the main beam with marginal effects on the flux 
scale accuracy. The main beam efficiency is 90\%. 

The flux calibration also accounts for the atmospheric opacity 
attenuation. The opacity, when observing the calibrator, was 0.085~dB at the observing Elevation (EL), that way the flux calibration applies a constant correction for 0.085~dB attenuation. During the observations the opacity ranged from 0.064 to 0.120~dB (including EL effects) for a maximum variation compared to the constant opacity correction of 0.035~dB (0.8\%), with marginal effects on the flux density scale accuracy. The opacity at the observing EL was computed from the zenithal opacity correcting for EL effects (1/cos(EL)), while zenithal opacity was computed from atmospheric parameters (temperature, pressure, and relative humidity).

\begin{figure}
\plotone{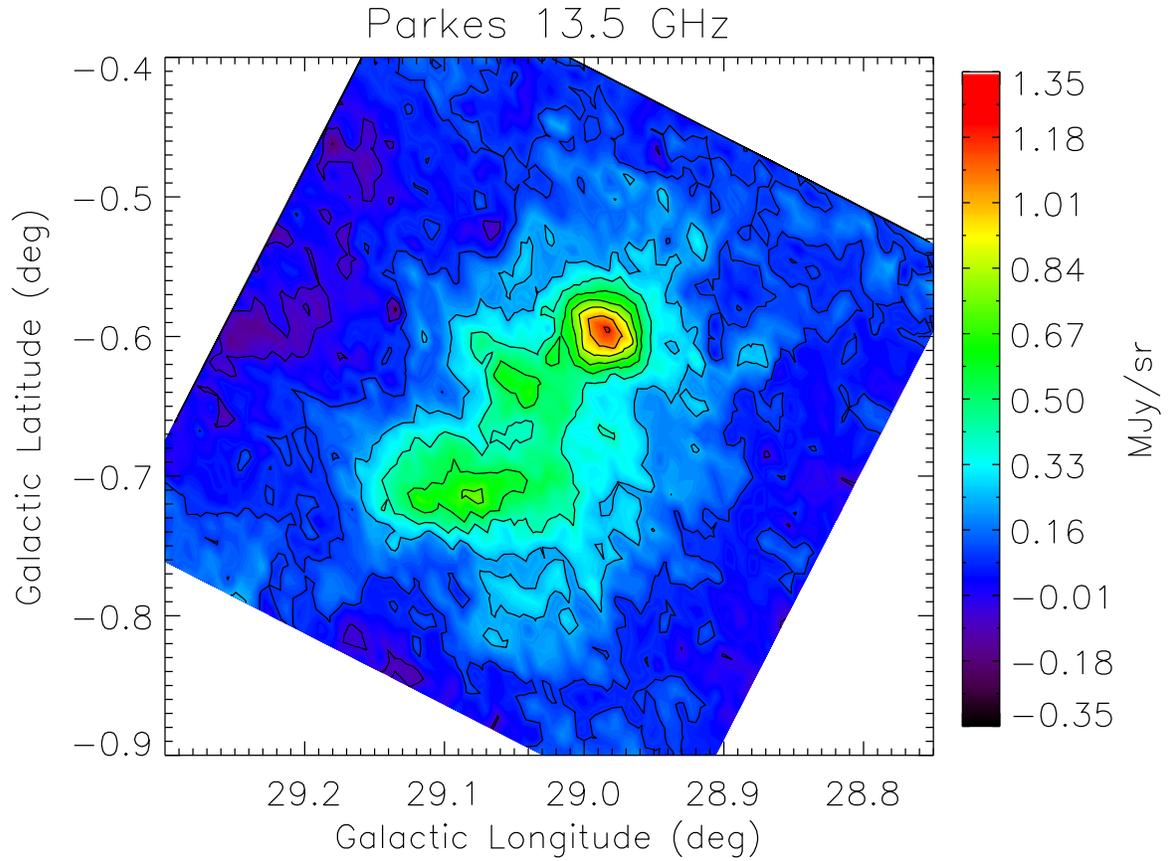}
\caption{13.5~{GHz} map of RCW175 obtained with the Parkes radio telescope using the Ku-band receiver. 
The angular resolution is 1.7~{arcmin} FWHM 
and sensitivity is 18~{mJy/beam}. Contours show levels each ranging  10\% of the peak emission. \label{map2}}
\end{figure}

The channels spanning the IF were then binned into seven 100-{MHz} sub-bands for the subsequent map-making
processing (see Figure \ref{map2}).  A standard basket-weaving technique with orthogonal scan sets along R.A. and 
Dec., spaced by 30~{arcsec}, was used to observe an area of $30'\times30'$ centered at the source. The scan speed 
and sampling time was the same as those for the 8.4~{GHz} observations. The same map-making software was applied to make 
the maps, and finally the seven sub-band maps were binned together into one. The final rms on the map is 
18~{mJy/beam} on beam-size scales, larger than the expected sensitivity (4~{mJy/beam}). Following 
the procedure described in Section~\ref{sec:8.4obs}, we estimate the Galactic emission contribution at 
{rms}$_{13.5\ \rm GHz}^{1.7'} = $13~{mJy/beam} at the frequency and resolution of the Ku-band 
observations, consistent with the observed {rms}.

\subsection{21.5~{GHz} observations}\label{sec:21.5obs}

The 21.5~{GHz} observations were conducted with the 13~{mm} receiver of the Parkes telescope on
 2011 August 30  for a total of 4~hr. The receiver covers a band of 16--26~{GHz}. We used the circular 
polarization setup with a band of 900~{MHz} centered at 21.55~{GHz}. The system temperature was 
$T_{\rm sys} \sim 95$~{K},  and the resolution FWHM = 67~arcsec. To detect the whole useful 
bandwidth the backend Digital Filter Banks Mark 3 (DFB3) was used with a configuration bandwidth of 
1024~{MHz} and 512 frequency channels~(2~MHz each). All autocorrelation and complex cross-products
 of the two circular polarizations were recorded ($RR^*$, $LL^*$, $RL^*$). The feed illuminates the primary 
mirror with an edge taper of 21~dB, the first sidelobe amplitude is $-30$~dB relative to the antenna pattern 
peak, and the gain is 1.70~{Jy}~{K}$^{-1}$. Observations consisted of 112 repeated scans of one strip along the major dimension of the source from 
({RA},{dec})=(281.50,$-3.82$)~{deg} to ({RA,dec})=(281.85,$-3.68$)~{deg} 
and back (see Figure \ref{regions}). This set-up was aimed at detecting polarized emission at low polarization 
fraction or to set a new relevant upper limit able to constrain the anomalous microwave emission model space. 

The flux density scale was calibrated using the source PKS B1921-293 with an assumed flux of 16.5~{Jy},
 and accuracy of 10\%. This is a variable source on a time scale of a few weeks and its flux density was measured 
with the Australia Telescope Compact Array the day after the Parkes observations on 2011 August 31.  As for 
the 8.4~{GHz} data, we could not measure the secondary lobes, and to estimate the error in flux accuracy 
for our extended source we estimated the antenna pattern following  \cite{gol87}. The source is mostly covered 
by the first three sidelobes, which extend to 12 {arcmin} (diameter). We found their combined 
solid angle to be 2.8\% of the main beam with marginal effects on the flux scale accuracy. The main beam efficiency 
is 88\%. 

As in the case of the 13.5~{GHz} observations, we have also included the atmospheric opacity attenuation in the flux calibration uncertainty. 
The opacity when observing the calibrator was 0.47~dB at the time and EL of the observation, and therefore the flux 
calibration applies a constant correction for 0.47~dB attenuation. 
During the observations the opacity ranged from 0.35 to 0.69~dB (including EL effects) 
for a maximum variation compared to the constant opacity correction applied 
of 0.22~dB (5\%). These, combined with the calibration uncertainty, degrade the 
total accuracy to 11\%.

On-axis instrumental polarization of the system over the observed band was 1.5\%. This was calibrated and 
corrected with observations of the planet Saturn. The procedure followed is standard: the fractional instrumental polarization on 
Stokes Q and U is measured as the fractional polarized response to an 
unpolarized source (Saturn in our case); then, for each piece of data to clean, 
the same fraction of Stokes I is subtracted from the measured Q and U. 
Estimate and correction have been done on each 2 MHz frequency 
channel. After correction, the residual instrumental polarization on 
our data set was 0.2\%. Off-axis instrumental polarization can be as high as 0.6\%. However, thanks to the on-axis Parkes configuration, to the particular low polarized and diffuse observed source, and to the fact that we have performed 1-dim scans with angles ranging from $60^{\circ}$ to $-20^{\circ}$ with respect to the horizon, cancellation effects apply to the residual contamination \cite[see e.g., ][]{car04,ode07}. Geometrical considerations accounting for the rotation of the observed field during the observational night, drive us to an estimate of a possible averaged residual polarization of 0.2\%. This (rather pessimistic) result drives our overall instrumental polarization uncertainty to 0.3\%.

The frequency channels over the 900~{MHz} IF band were binned in 90 sub-bands for flux and instrumental 
polarization calibration. All of the sub-bands were then combined into one for the subsequent analysis. Scan length, 
rate, and sampling time was as for the observations at 8.4~{GHz}. We reached a sensitivity per beam-sized 
pixel of $\sigma_{Q,U}^{21.5\ \rm{GHz}}$~=~0.2 {mJy/beam} in polarization, consistent with the expected
 value.  The fluctuations in Stokes I are larger with an {rms} of 5.4~{mJy/beam}. This is 
consistent with the {rms} of the Galactic signal that, following the same procedure of Section~\ref{sec:8.4obs}, 
we can estimate at {rms}$_{21.5\ GHz}^{67"'} = $5.6~{mJy/beam} at the frequency and resolution of 
these observations. 

It is worth noticing that the rms larger than the expected statistical noise is well 
accounted by the Galactic emission fluctuations at all three frequencies. 
Another possible source of departure from the expected statistical rms is the 1/f noise. This is drastically mitigated by the map making procedure and the behaviour of the additional noise we measure (decreasing with frequency) is opposite to what we would expect from this type of noise. This, along with the Galactic emission accounting for the measured rms, makes a possible contribution of the 1/f noise a minor term of the error budget of our observations.

\section{Spectral Energy Density}\label{sed}

\subsection{Radio and microwave ancillary data}

Publicly available ancillary data have been analyzed and combined with our Parkes observations to derive the SED of RCW175. 
At 1.4~{GHz} we have considered the NRAO VLA Sky Survey (NVSS) \citep{con98} data, characterized 
by a relatively high angular resolution (FWHM = 0.75~{arcmin}), although they lack sensitivity on scales 
larger than 15~{arcmin}. Effelsberg 1.4~{GHz} data by \cite{rei90a} are also available\footnote{http://www.mpifr-bonn.mpg.de/en/effelsberg}, but these 
need to be re-calibrated through the Stockert 1.4~{GHz} data \citep{rei82}, which are characterized by 
too low a resolution (34~{arcmin} FWHM) for our analysis. We have thus analyzed and used, to constrain the integrated 
1.4~{GHz} emission, the Green Bank 300~{ft} data by \cite{alt70}, which are affected by a larger 
uncertainty but better match our angular resolution and map size (FWHM $\approx$ 10~{arcmin}). 
Data from the Effelsberg 2.7~{GHz} \citep{rei90b} and Parkes 5~{GHz} \citep{hay78} surveys match the angular resolution needed 
for the present analysis and thus we have included the flux density estimated by \cite{tib12} for these surveys.

In addition to these data, we performed an integrated flux density analysis on the GBT data (8.35~{GHz}, and 14.5~{GHz}) 
by \cite{lan00}, as well as the Nobeyama 10~{GHz} data by \cite{han87}, which have been considered in
 the analysis of the integrated flux density for the RCW175 HII region. These data, however, are affected by significant 
systematics on angular scales that are important for our study. For instance, the Nobeyama data present a non-constant offset 
which is probably due to an inaccurate removal of the Galactic plane emission. For this reason, they are not included in
our analysis. At microwave frequencies, as well as for our new Parkes data, we have also used the  CBI 31~{GHz} \citep{dic09}
and VSA 33~{GHz} \citep{tod10} flux density estimates. A summary of the data-set used in this work is provided in 
Table~\ref{tab:seddata}.

\subsection{{Millimetric}, {sub-millimetric}, and {IR} ancillary data}

We have analyzed the {mm}, {sub-mm}, and {IR} emission of RCW175 using data from the 
\Planck~satellite, the \Herschel~Space Observatory, the InfraRed Astronomical Satellite (\IRAS), and the 
\Spitzer~Space Telescope. The first release of the \Planck~maps \citep{pla11b} has been available since April 
2013 and includes maps in nine frequency channels, ranging from 30 to 857~{GHz}. An overview of Planck products can be found in \cite{pla14b}. CO-corrected \Planck~data \citep{pla14c} have 
allowed us to place upper limits or obtain marginal detections at millimetric wavelengths from 100 to 353~{GHz}. 
For the color corrections we have integrated the fitting model in the \Planck/HFI bandpasses, applied the 
corrections, and iterated until convergence, obtaining corrections ranging from 3\% to 10\% from 100 to 353~{GHz}. 
For the \Herschel~data, we have used the maps from the Hi-GAL project, a \Herschel~Open Time Key Programme~\citep{mol10}.
Hi-GAL maps were obtained by PACS and SPIRE parallel mode observations,  and cover five bands between 545 and 
4300~{GHz}. The maps have been created using a dedicated map-making algorithm \citep{traf10,piazzo12}, 
optimized for regions with high contrast, such as the Galactic plane. For \IRAS, we have used the 60 $\mu$m and 
100 $\mu$m  IRIS (\IRAS~reprocessed, \cite{mamd05}) maps. Finally, we have used the \Spitzer~MIPS 24~$\mu$m  
data from the MIPSGAL survey \citep{carey09}. A summary of the characteristics of the {mm}, {sub-mm}, 
and {IR} data used in this work is given in Table~\ref{tab:seddata}.

\begin{table}
\caption{Integrated fluxes and instrumental properties used for the RCW175 analysis and Spectral Energy Density fit. The fluxes have been estimated through aperture photometry of the available data (see text for details). 
Uncertainties are obtained by quadrature summation of the statistical and systematic/calibration uncertainties.\label{tbl-2}}                       
\label{tab:seddata}
\footnotesize
\begin{tabular}{l l l c l l}
\hline\hline
Frequency & Telescope /  &  Ang. res.  &  Flux Density  & Calibration & Reference  \\
(GHz) &  Experiment  &  FWHM &  (Jy) & uncertainty &   \\
\hline
1.4  & Green Bank 300~ft  & ~9$^\prime$.4$\times$10$^\prime$.4  & $6.0\pm1.8$ & included & \cite{alt70}  \\
2.7  & Effelsberb 100~m  & ~4$^\prime$.3  & $5.7\pm0.9$ & included & \cite{rei90b}  \\
5.0  & Parkes 64~m   & ~4$^\prime$.1 & $4.0\pm0.8$ & included & \cite{hay78}  \\
8.4  & Parkes 64~m  & ~2$^\prime$.4 & $4.86\pm0.51$ &  $5\%$  & This work  \\
13.5  & Parkes 64~m   & ~1$^\prime$.7 & $5.71\pm0.62$ &  $5\%$  & This work  \\
31  & CBI  & ~4$^\prime$.3$\times$4$^\prime$.0  & $5.97\pm0.30$ & $1.3\%$ & \cite{dic09}  \\
33 & VSA  & ~13$^\prime$.1$\times$10$^\prime$.0  & $7.8\pm1.6$  & $\lesssim 2\%$ & \cite{tod10}  \\
100  & \Planck~HFI  & ~9$^\prime$.59  & $3.2\pm2.7$ & 0.5\% & \cite{pla11b}  \\
143  & \Planck~HFI  & ~7$^\prime$.18   & $3.1\pm3.5$ & 0.5\% & \cite{pla11b}  \\
217  & \Planck~HFI  & ~4$^\prime$.81  & $23\pm22$  & 0.5\% & \cite{pla11b}  \\
353  & \Planck~HFI  & ~4$^\prime$.70  & $81\pm92$  & 1.2\% & \cite{pla11b}  \\
600  & \Herschel~SPIRE-500  & ~36$^{\prime\prime}$.3 & $650\pm130$    & 7\% & \cite{spire11}  \\
857  & \Herschel~SPIRE-350 & ~24$^{\prime\prime}$.9 & $1580\pm320$    & 7\% & \cite{spire11}  \\
1200  & \Herschel~SPIRE-250  & ~18$^{\prime\prime}$.2  & $4110\pm830$    & 7\% & \cite{spire11}  \\
1875  & \Herschel~PACS-160  & ~10$^{\prime\prime}$.35  & $9660\pm1940$    & 7\% & \cite{balog13}  \\
3000  & \IRAS~IRIS-100  & ~4$^\prime$.3  & $10300\pm1800$    & 13.5\% & \cite{mamd05}  \\
4286  & \Herschel~PACS-70  & ~8$^{\prime\prime}$.0  & $9600\pm1900$    & 7\% & \cite{balog13}  \\
5000  & \IRAS~IRIS-60  & ~4$^\prime$.0  & $6740\pm1050$    & 10.4\% & \cite{mamd05}  \\
12670  & \Spitzer~MIPS-24  & ~6$^{\prime\prime}$.0  &     & 10\% & \cite{carey09} \\
\hline
\end{tabular}
\end{table}

\subsection{Integrated emission spectrum and fit}

\begin{figure}
\epsscale{1.00}
\plotone{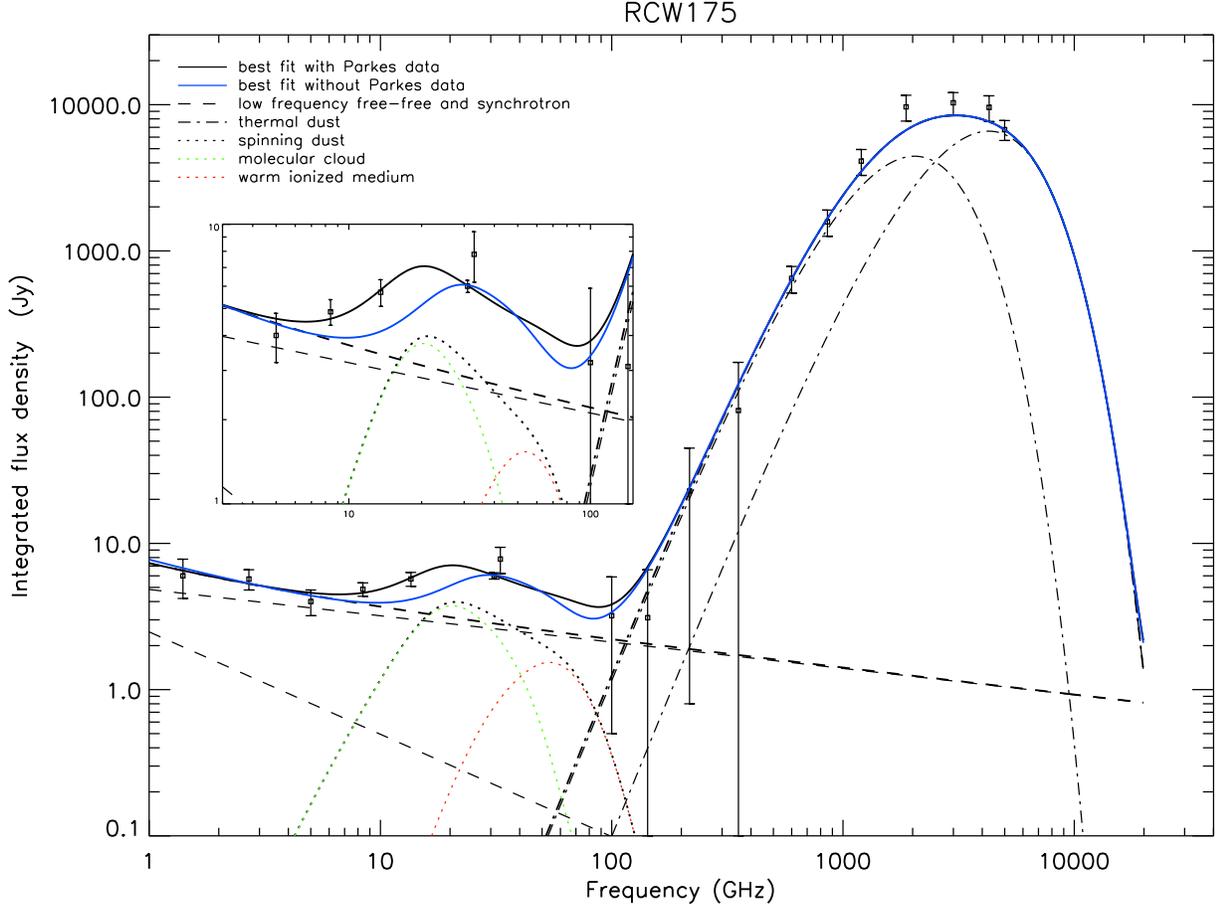}
\caption{SED of RCW175 obtained with aperture photometry in a 12~{arcmin} radius. The fit is performed 
with free--free, synchrotron, thermal dust emission,  and different spinning dust models. The black solid line 
denotes the sum of all the fitted components. Dashed lines indicate the low frequency ({i.e.,} free--free
 and synchrotron) components, dot-dashed lines show the thermal dust fit, and dotted lines denote the spinning 
dust models. The blue line shows the same fit, excluding the Parkes data. The overall fit suggests the  presence
of at least two different components of anomalous emission. See the text and the figure legend for more details. \label{sed_plot}}
\end{figure}

The integrated emission from the RCW175 HII region, presented in Figure \ref{sed_plot}, was obtained 
using aperture photometry  with an annular radius of 12~{arcmin} and an estimate of background 
in the annular region from the 12--17~{arcmin}  range. Uncertainties were estimated from the
 map fluctuations outside the source. This procedure allows us to neglect structures larger than  
$\sim$ 15 {arcmin}. Using the data set described above, we have fitted a model that includes 
the four emission mechanisms that we have assumed  to be dominant in the microwave range for RCW175: 
synchrotron, free--free, spinning dust, and thermal dust emission.

The thermal dust was fitted to a minimum wavelength of 60$\mu$m, using the IR, {sub-mm}, and
 {mm}  data and a sum of two modified black-bodies, $S_{\nu} \propto \nu^{\beta}BB(\nu,T_{\rm dust})$,
 as in \cite{tib12}.  In this fit, upper limits or tentative detections derived in the Planck 143, 217, and 
353~{GHz} bands \citep {pla11b} have also been  included. The Planck 100~{GHz} data point was 
instead included in the ``low frequency'' fit,   in order to account for the impact of the thermal dust 
Rayleigh--Jeans tail. Planck data at frequencies above 353~{GHz} were not considered, as we rely on the 
higher spatial resolution SPIRE data. From the fit, we retrieve results consistent with \cite{tib12}, 
{i.e.,} a cold dust population with temperature $T_{\rm c}=24.3^{+2.4}_{-3.2}$ K, and a warm dust population
 with  temperature $T_{\rm w}=50.8^{+1.5}_{-1.4}$ K, having assumed a dust spectral emissivity index $\beta=2$ for both populations. 
The interstellar radiation field~(ISRF) plays an important role in heating and exciting the smallest dust grains and PAHs. 
This can be quantified though the parameter $\chi_{\rm ERF}$ which can be estimated as 
$\chi_{\rm ERF}=(T_{\rm c}/17.5K)^{(4+\beta)}$. We find $\chi_{\rm ERF}=7$, which is indicative of a radiation field 
intensity a few times the ISRF in the solar neighborhood \citep{mat83,ali09,pla14a}. The dust temperatures 
are slightly higher with respect to most of the AME sources, which are found to be associated mostly with 
colder regions \citep{pla14a}; however, they follow the same trend as found by \cite{tib11} for the Perseus complex, 
where the strength of the ISRF was found to be of key importance for AME regions. 

The low-frequency data ($<$ 8~{GHz}) were initially fitted with a single power-law ({i.e.,} 
$S_{\nu} \propto \nu^{\alpha}$), which resulted in an average spectral index of $\alpha=-0.28^{+0.09}_{-0.06}$.
Later, a fit was performed by accounting for a free--free component plus a synchrotron component, with intensity 
and spectral index left free to vary ($S_{\nu} \propto A_{\rm ff}\nu^{\alpha_{\rm ff}} + A_{\rm sy}\nu^{\alpha_{\rm sy}}$). The 
results show that, below 8~{GHz}, the emission is dominated by free--free emission with spectral index 
$\alpha_{\rm ff}=-0.18 \pm 0.07$, with an additional sub-dominant contribution due to a steeper power-law with spectral 
index $\alpha_{\rm sy}=-0.7^{+0.5}_{-0.2}$, consistent with synchrotron emission. At 1~{GHz}, the fraction of 
free--free emission with respect to synchrotron emission is $A_{\rm ff}/A_{\rm sy}=1.8$, clearly increasing with frequency
 and indeed  confirming the hypothesis of synchrotron contamination along the line of sight of RCW175 as previously found by \cite{tib12}. To fit the 
spinning dust component, we used the \textsc{spdust.2.01}\footnote{http://www.sns.ias.edu/$\sim$yacine/spdust/spdust.html} code \citep{ali09,sil11} to compute different 
emission spectra accounting for the radiation field intensity and gas temperature of this region. The hydrogen 
number density and gas temperature have been chosen to be consistent with standard values associated with Molecular
Clouds (MC) and the Warm Ionized Medium (WIM) \citep{dra98, dic12}. For the WIM component we have assumed 
hydrogen number densities $n_{\rm H}^{\rm WIM}=26.3$/cm$^{3}$ derived from the 5~{GHz} integrated flux of RCW175 
and assuming an electron temperature, $T_{\rm e}=5800$ K, following the analysis presented in \cite{pal04} and \cite{tib12}. As for the MC 
component, we built a grid of spectra with $n_{\rm H}^{\rm MC}$ varying from $n_{\rm H}^{\rm MC}=100$/cm$^{3}$ to 
$n_{\rm H}^{\rm MC}=1000$/cm$^{3}$, with a stepsize of 25/cm$^{3}$, and performed a best fit over $n_{\rm H}^{\rm MC}$. The best fit was 
obtained for $n_{\rm H}^{\rm MC}=150$~/cm$^{3}$, corresponding to a minimum $\chi^{2}$/d.o.f. = 0.78. The major ion 
fractions, [H+] and [C+], are estimated as described in \cite{ysa11}. Fits have been performed using \textsc{MPFIT} 
\citep{mar09} with physically reasonable priors, such as ensuring non-negative values. Attention has also been paid to ensure that the fitted parameters
 avoid hitting the prior limits. The parameter uncertainties were determined through
 Monte Carlo analysis by randomising the data points by their 1-$\sigma$ uncertainty and evaluating the scatter as the final uncertainty. In Table \ref{fit} we collect the parameters resulting from the aforementioned fits.

\begin{table}
\begin{center}
\caption{Best-fit parameters obtained form the SED fit of the integrated emission of the RCW175 HII region. 
The two dust temperatures, $T_{\rm c}$ and $T_{\rm w}$, were obtained from the fit of the sum of two modified
 black-bodies. The dust emissivity $\beta$ has been taken equal to 2. The ISRF has been parameterized through 
$\chi_{\rm ERF}$.  The low frequency fit was either performed through a single power-law with spectral index 
$\alpha$, or through the sum of two power-laws of spectral indices $\alpha_{\rm ff}$ and $\alpha_{\rm sy}$, and 
normalization $A_{\rm ff}$  and $A_{\rm sy}$, respectively.  The spinning dust models are constrained with the hydrogen 
number densities $n_{\rm H}$ and the column density $N_{\rm H}$ for an MC and a WIM component. \label{fit}}
\footnotesize
\begin{tabular}{lll}
\tableline\tableline
Parameter & Value & Notes \\
\tableline
      $T_{c}$                &  $24.3^{+2.4}_{-3.2} ~K$                                    & Sum of two modified black-bodies \\
      $T_{w}$               & $50.8^{+1.7}_{-1.8} ~K$                                     & Sum of two modified black-bodies  \\
      $\beta$                &  $2$                                                                    & Fixed  \\
      $\chi_{ERF}$               & $7$                                                             & Derived from $T_{c}$ \\
      $\alpha$              & $-0.28^{+0.09}_{-0.06}$                                     & Single power-law\\
      $A_{ff}$               & $4.8^{+0.9}_{-1.0}~Jy$                                      & Sum of two power-laws \\
      $\alpha_{ff}$       & $-0.18 \pm 0.07$                                                 & Sum of two power-laws\\
      $A_{sy}$             &  $2.5^{+0.7}_{-0.7}~Jy$                                      &  Sum of two power-laws\\
      $\alpha_{sy}$      & $-0.7^{+0.5}_{-0.2}$                                           & Sum of two power-laws\\
      $T_{e} $              &  $5800 ~K$                                                         & Derived from 5~GHz integrated flux \\
      $n_{H}^{WIM}$   &  $26.3~/cm^{3}$                                                  & Derived from 5~GHz integrated flux  \\
      $N_{H}^{WIM}$  & $(0.057^{+0.004}_{-0.022})\times10^{22} ~H/cm^{2}$ &  \\
      $n_{H}^{MC}$     & $150~/cm^{3}$                                      & Range$=[100/cm^{3}-1000~/cm^{3}]$; step$=25~/cm^{3}$\\
      $N_{H}^{MC}$    & $(3.8^{+0.4}_{-0.8})\times10^{22} ~H/cm^{2}$             & \\
      \tableline
\end{tabular}
\end{center}
\end{table}

\subsection{SED interpretation}

The total integrated flux SED confirms and strengthens the presence of AME in the HII region RCW175. The 
contribution from UCHII regions in RCW175 was found to be negligible by \cite{tib12}. We note that 
repeating the low frequency fit both including and excluding the Parkes data (which fill the gap between the 
radio and {centimetric} data) provides useful information. Additional useful information could be obtained 
from $90-100$~GHz high angular resolution data, as these could break a possible degeneracy between AME and
a strong thermal dust Rayleigh--Jeans tail at {millimetric} wavelengths. By excluding the Parkes data, the 
SED is well fitted by a single cold spinning dust component (see the blue curve in Figure \ref{sed_plot}). On
 the other hand, including the Parkes data favors a model with an additional spinning dust component consistent with the WIM, similarly to what was presented by \cite{pla11a}.
In the case of our measurements, this is mainly due to the fact that the rise in the SED in the microwave range  ({i.e.,} 
$\sim5-30$~{GHz}) occurs at lower frequencies with respect to what a single component with physically 
meaningful characteristics allows. As shown by \cite{ali09} and \cite{ysa11}, the peak frequency of the 
spinning dust emission is mildly sensitive to the radiation field and the gas temperature, and highly sensitive 
to the total hydrogen number density $n_{\rm H}$. The lower $n_{\rm H}$, the lower the peak frequency, resulting in 
the necessity of adding a second component with low $n_{\rm H}$ to better fit the data. We should, however, stress
 that the modeled anomalous microwave emission SED is strongly model dependent, with significant degeneracy between 
some of the parameters. We consider this result as an example of the capability of new data to break parameter 
degeneracies: more data, sampling the microwave frequency space at higher resolution, are necessary to solve this
 degeneracy, together with more detailed simulations and modeling. Nevertheless, the Parkes data we use here 
allow us to make a first attempt in this direction and clearly show that one spinning dust component is
not sufficient to reproduce the observations. Combining different hydrogen number densities $n_{\rm H}$, and fitting 
for the column density $N_{\rm H}$, we find a best fit for a superposition of  WIM and  MC components with
 $N_{\rm H}^{\rm WIM}=(0.057^{+0.004}_{-0.022})\times10^{22}$ H/cm$^{2}$, and $N_{\rm H}^{\rm MC}=(3.9^{+0.4}_{-0.8})\times10^{22}$ H/cm$^{2}$, 
and with hydrogen number densities $n_{\rm H}^{\rm WIM}=26.3$/cm$^{3}$ (fixed) and $n_{\rm H}^{\rm MC}=150 \pm 25$(step) /cm$^{3}$, consistent with the values found by \cite{tib12}. This suggests that the RCW175 emission results from different components distributed across the region. In particular, 
we can speculate that the WIM component is associated with the interior of the HII region, and the MC component
 with the surrounding photo-dissociation region (PDR). 
 
From the ratios $N_{\rm H}/n_{\rm H}$ we derive lengths along the line of sight of the structure producing spinning dust emission of $7.1$ and $84.3$ pc, respectively for the WIM and the MC components. Considering the distance to RCW175, $3.2\times 10^3$~pc \citep{tib12}, and the transversal size derived from our images, $\sim$20~arcmin, we estimate a transversal physical size of $18.6$~pc. This is of the same order as the line-of-sight length of the WIM component. On the other hand, the corresponding length of the MC component is slightly higher than what one would expect under the assumption of the structure having similar sizes along the line of sight and on the plane of the sky.

\section{Morphology}\label{morphology}

\begin{figure}
\epsscale{1.00}
\plotone{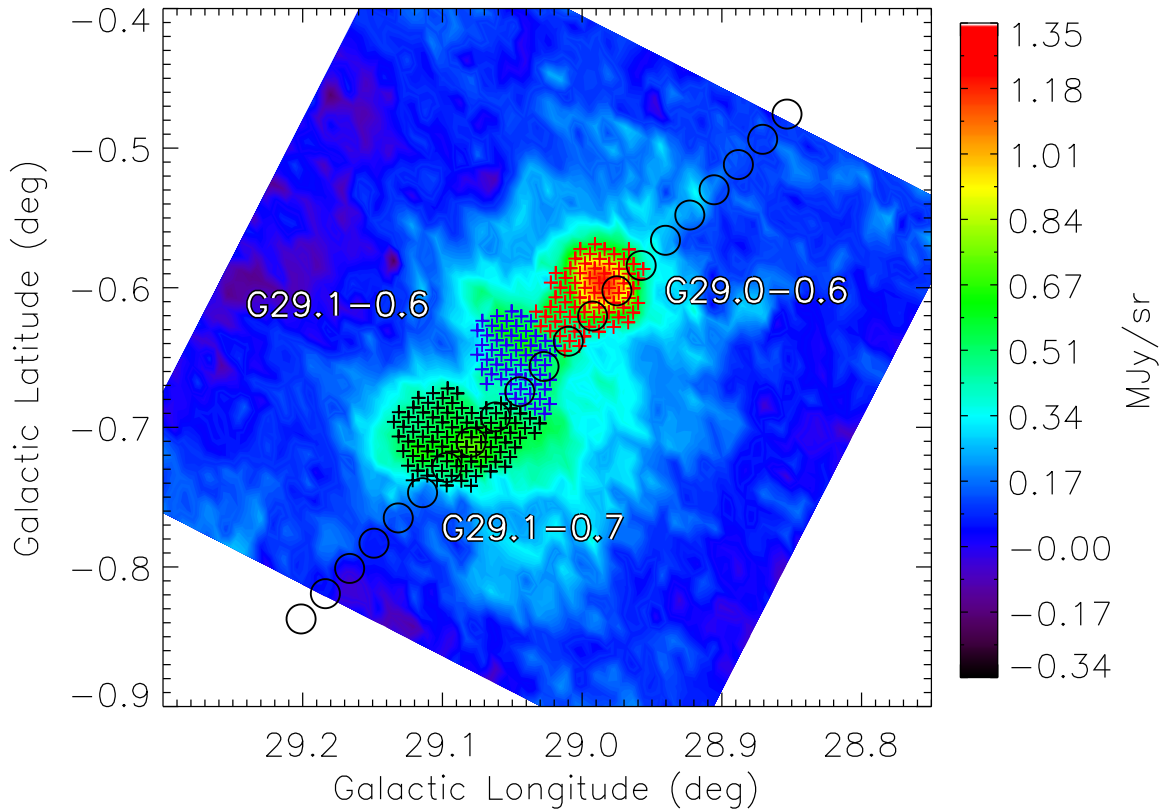}
\caption{Parkes 13.5~{GHz} map of RCW175 with the three identified sky regions highlighted. We performed 
our cross-correlation analysis, presented in section \ref{ir}, to the regions covered by crosses: red crosses 
are used to denote the compact and bright G29.0$-$0.6 region, blue crosses identify the G29.1$-$0.6 central region, 
and black crosses refer to G29.1$-$0.7.
The polarization observation scan is also indicated with circles representing the size of the beam FWHM at 
21.5~{GHz}. \label{regions}}
\end{figure}

The morphology of the RCW175 HII region has been studied by \cite{dic09} and others, who have identified two sub-regions
 within RCW175 (G29.0$-$0.6 and G29.1$-$0.7) and built the spectrum of the whole region, as well as the spectrum of its 
brightest constituent (G29.0$-$0.6). \cite{tib12} performed a detailed multi-wavelength study of RCW175, and described
 G29.0$-$0.6 as the brighter, more dusty component, with G29.1$-$0.7 being described as the more diffuse, more evolved
 component. So far, the only high angular resolution (that is better than 2~{arcmin}) maps of RCW175 have 
been the 1.4~{GHz} NVSS maps \cite[{i.e.,} NVSS data at 1.4~{GHz} by][]{con98}, at radio 
frequencies and, in the infrared, the \Spitzer~and \Herschel~maps. Our new microwave maps allow a more detailed study 
of the different spatial components of this HII region also in this frequency domain. In our maps, we can clearly 
identify the brighter and unresolved G29.0$-$0.6 region toward the West. In addition, we resolve the more diffuse G29.1
region into two separate components, {i.e.,} G29.1$-$0.7 towards the east, and G29.1$-$0.6 towards the north (see 
Figure \ref{regions}). A similar structure can be seen in the 1.4~{GHz} NVSS data \citep{con98} and in the
10~{GHz} Nobeyama map \citep{han87}. As mentioned, with both of these data sets, we could not extract 
information about the larger angular scales within the region. Despite the limitations, we could probe the self-similarity 
of the spectral behavior of each region.

Relying on the quality and on the good (or common) systematic effect control of the Parkes maps, as well as on
 the self-similarity of the SED of the different constituents of the region studied, we have attempted to perform a 
simple component separation analysis. 

\begin{figure}
\epsscale{1.00}
\plotone{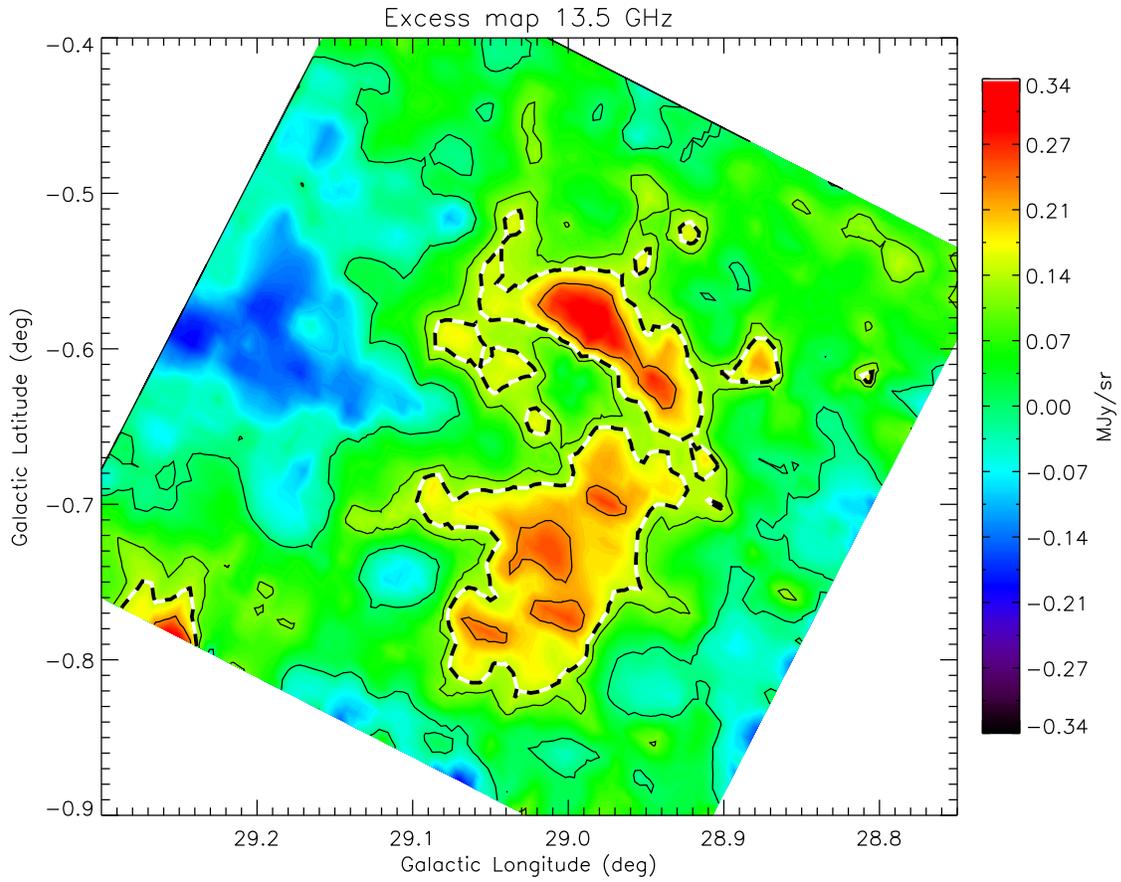}
\caption{RCW175 map at 13.5~{GHz} displaying the emission exceeding the expected level of synchrotron 
and free--free emission. Contours show levels separated as in Figure \ref{map}. Dashed line indicates the level
 corresponding to 1-$\sigma$ uncertainty (see text for details). \label{gfit3}}
\end{figure}

To do this, we first smoothed the 13.5~{GHz} map to the same angular resolution as the 8.4~{GHz} map 
({i.e.,} FWHM = 2.4~{arcmin}). We then extrapolated the 8.4~{GHz} map to 13.5~{GHz}, assuming
an average spectral index power-law of $\alpha$ = $-0.28$ (as derived in 
Section 3), and subtracted it from the actual 13.5~{GHz} map, producing a map of the AME at 13.5~{GHz} (see Figure \ref{gfit3}).
For this analysis we have assumed that the 8.4~{GHz} map is not contaminated by AME, resulting in a bias in the 
extracted map, which should thus be treated as an underestimate of the AME in the region. 
The difference map clearly reveals residual emission which is mainly concentrated in the compact
component, G29.0$-$0.6, plus an additional diffuse component surrounding the G29.1 complex. In Figure \ref{gfit3} we use a dashed line to illustrate 
the level corresponding to 1-$\sigma$ uncertainty. The internal part thus exceeds the 
1-$\sigma$ uncertainty where we have included a 5\% error arising from the modeling uncertainty.  We have also included 
in the uncertainty the propagated error arising from the fit affecting the estimation of the flux. 
We emphasize that this is one of the first attempts to perform a detailed component separation at microwave wavelengths on a single 
galactic source of AME. This approach will likely represent a pathfinder for detailed studies of the physics of the individual ISM components.

\section{IR data}\label{ir}

\begin{figure}
\epsscale{1.0}
\plotone{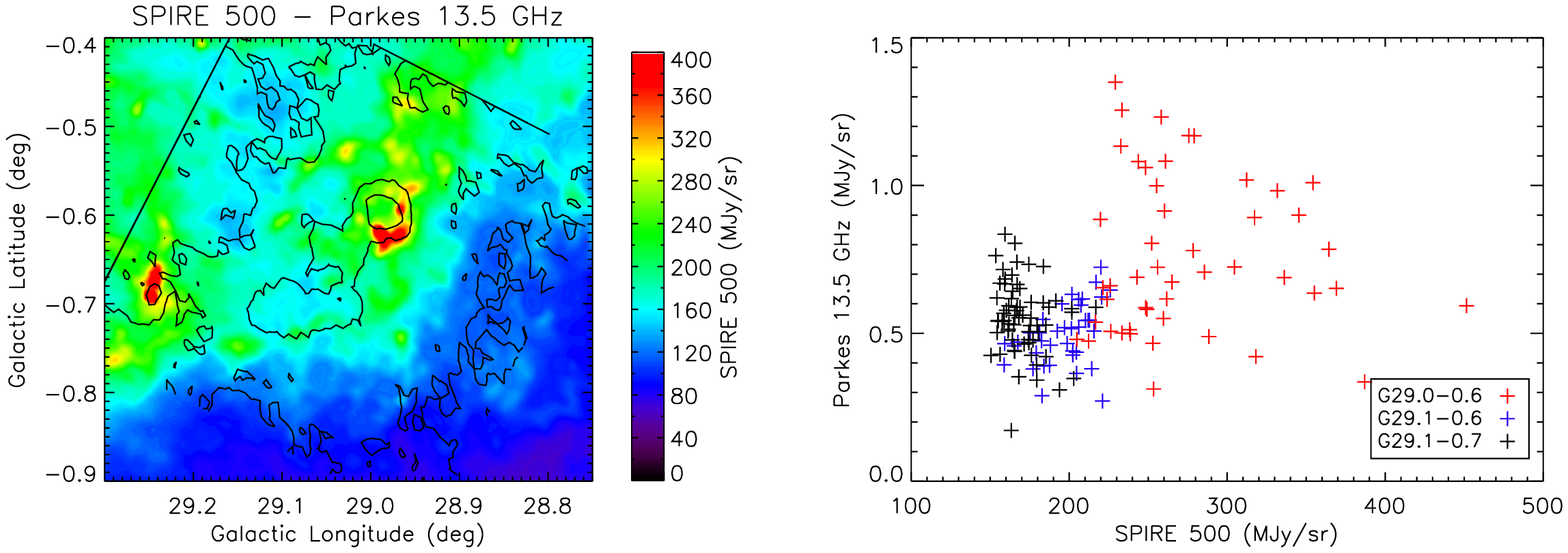}
\plotone{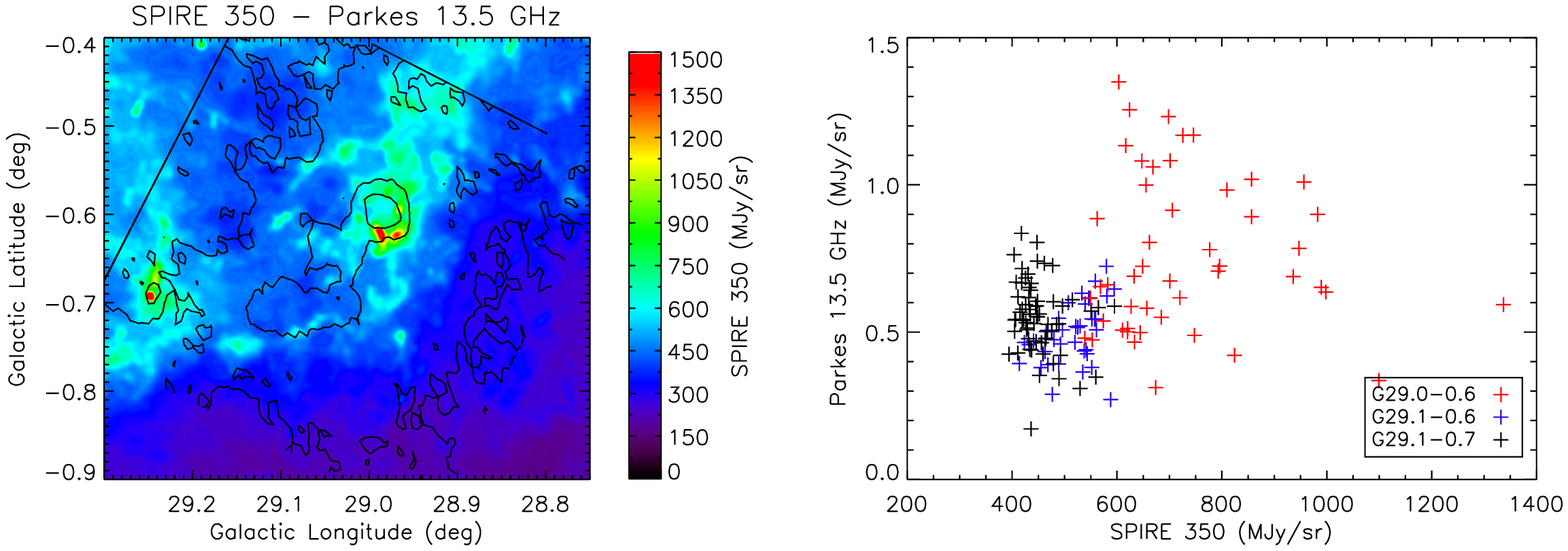}
\plotone{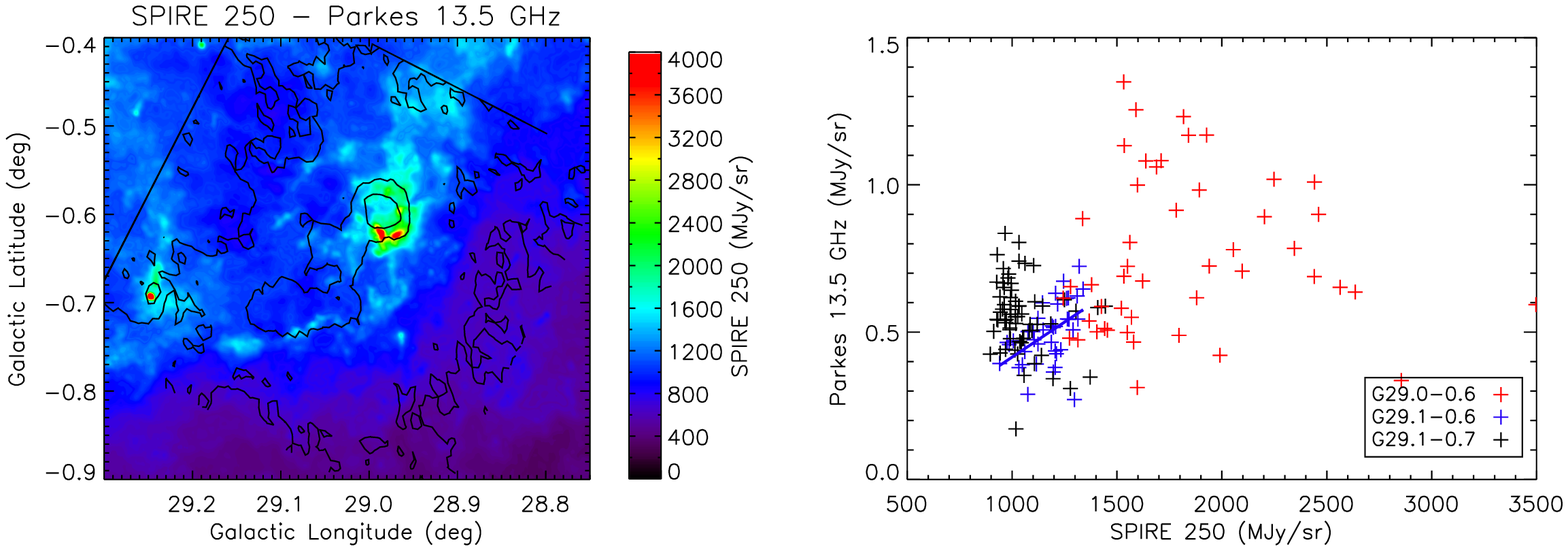}
\caption{From top to bottom: Ku-band 13.5~{GHz} maps vs SPIRE-500~$\mu$m, SPIRE-350~$\mu$m, and 
SPIRE-250~$\mu$m maps. The {IR} maps, with the 13.5~{GHz} in contours (with levels at 0\%, 33\% and 66\% of the peak values), are shown on the left. 
The corresponding scatter plots, {i.e.,} {microwave} map vs.\ {IR} map, are shown on the right.
 Different colors identify different regions, following the color scheme in Figure~\ref{regions}.}
\label{fig:ttplot13vs500_350_250}
\end{figure}

\begin{figure}
\epsscale{1.0}
\plotone{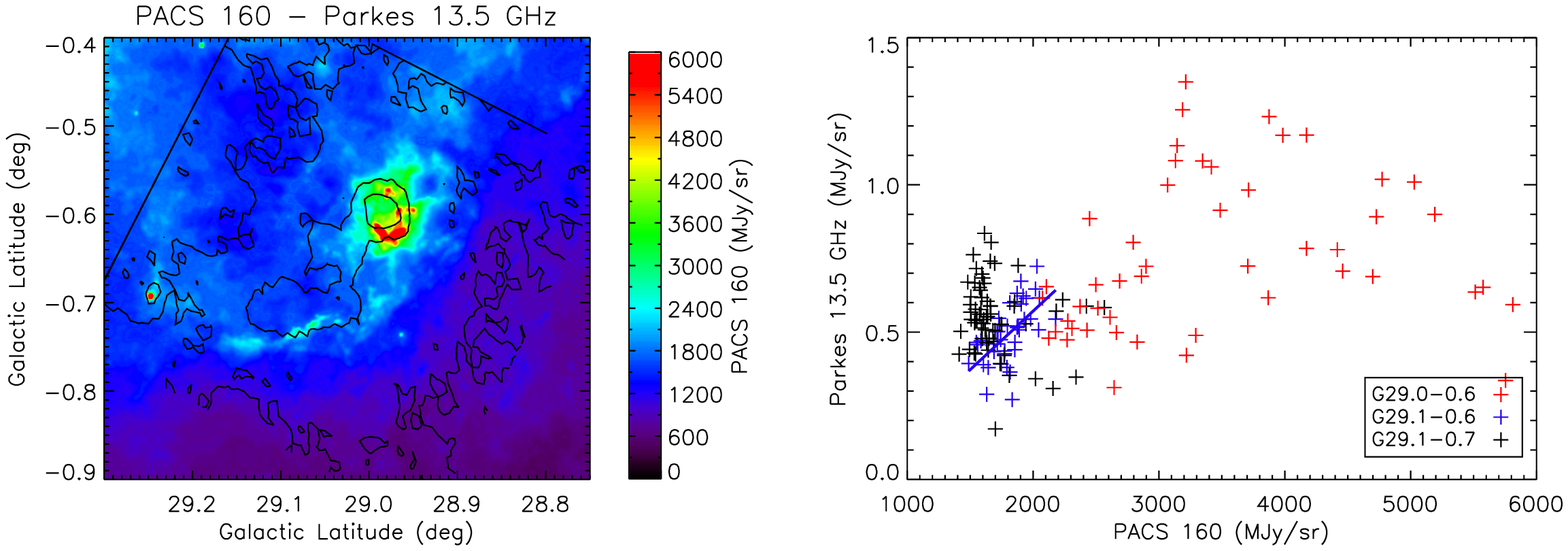}
\plotone{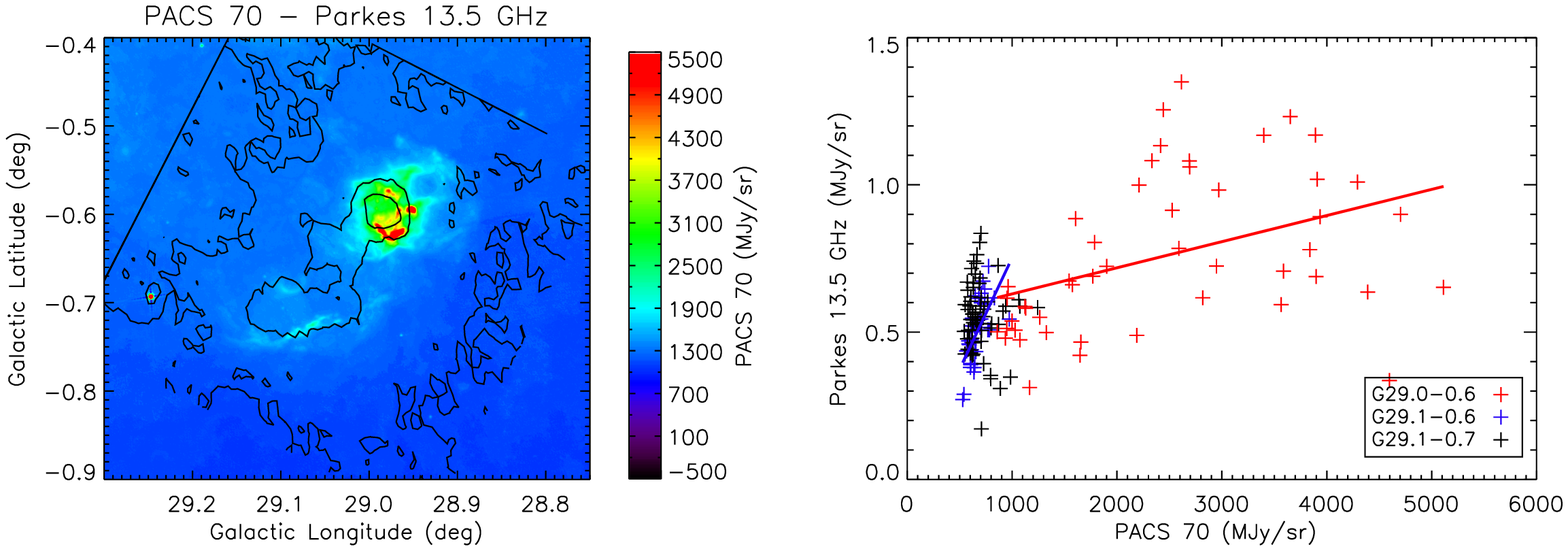}
\plotone{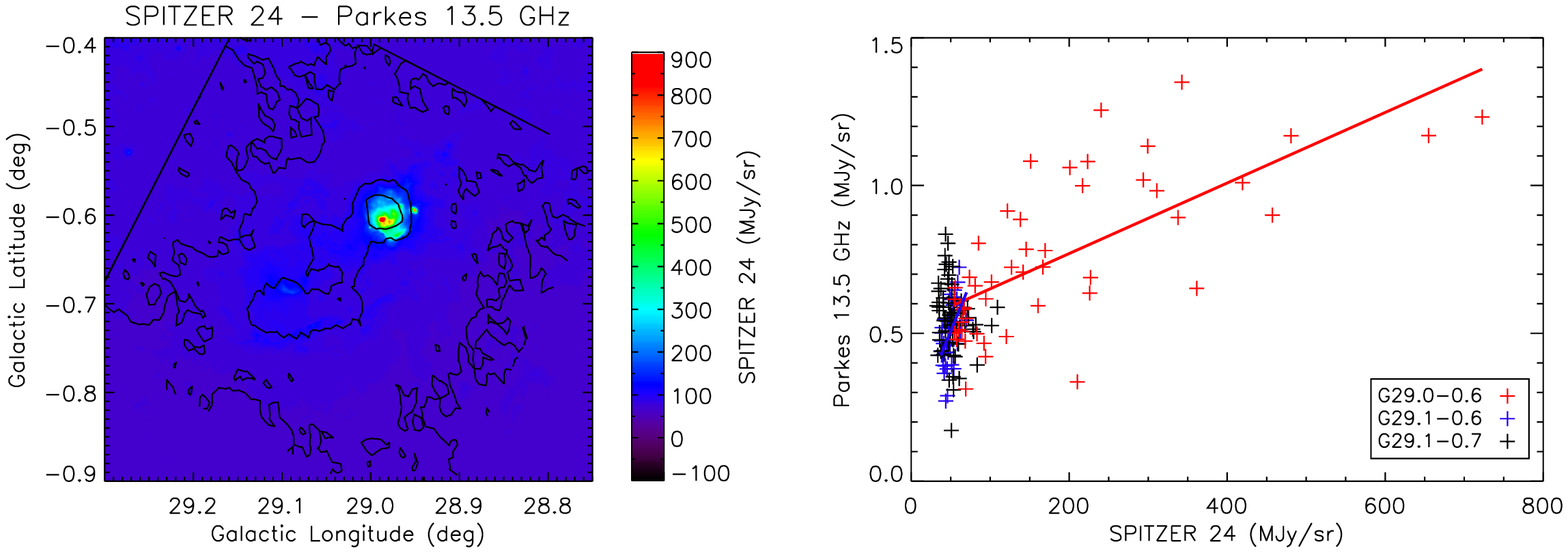}
\caption{From top to bottom: Ku-band 13.5~{GHz} map vs.\ PACS-160~$\mu$m, PACS-70~$\mu$m, and 
MIPS-24~$\mu$m maps. The corresponding scatter plots, {i.e.,} {microwave} map vs.\ {IR} 
map, are shown on the right. Different colors identify different regions, following the color scheme in Figure~\ref{regions}.} 
\label{fig:ttplot13vs160_70_24}
\end{figure}

A cross-correlation analysis between radio-frequency and IR data can improve our understanding of the physical 
components of the region. In Figure \ref{fig:ttplot13vs500_350_250}, and \ref{fig:ttplot13vs160_70_24} we show the correlation analysis 
we performed using the Ku-band 13.5~{GHz} data and the SPIRE, PACS, and MIPS IR data, which span the 
wavelength range from 500~$\mu$m to 24~$\mu$m. To this end, we have convolved all the maps to the 13.5~{GHz} 
data angular resolution. On the left, we show the {IR} maps and, as contours, the map at 
13.5~{GHz}; on the right, we provide the corresponding scatter plot, {i.e.,} the {IR} map vs.\   
the {microwave} map, where different regions in the map are marked with different colors, following the same color 
scheme defined in Figure~\ref{regions}. Red crosses are used to denote the compact and bright G29.0$-$0.6 region, blue crosses 
identify the G29.1$-$0.6 central region, and black crosses refer to G29.1$-$0.7. 

Noteworthy, the combination of {sub-mm} and IR bands is typically used to trace the three major populations of dust 
grains found in the ISM. In the reference framework of dust emission in the diffuse ISM ($T_{\rm d} \sim$ 17 K), the 
SPIRE (500~$\mu$m, 350~$\mu$m and 250~$\mu$m), and PACS 160~$\mu$m bands trace thermal dust emission from Big 
Grains, while the PACS 70~$\mu$m and MIPS 24~$\mu$m bands trace emission from Very Small Grains (VSGs). This 
picture still holds (apart for the 24~$\mu$m emission that we discuss below) in the case of HII regions. 

From the 13.5~{GHz} vs SPIRE-500~$\mu$m correlation, we note an absence of correlation for G29.1$-$0.6 and 
G29.1$-$0.7,  and a mild correlation, although with a large scatter, for G29.0$-$0.6. A more detailed analysis reveals 
a shell-like structure from which the {IR} emission probably originates, while the microwave emission appears 
to generate from the bulk of G29.0$-$06. A similar behavior is highlighted by the correlation with the 350~$\mu$m, 
250~$\mu$m, 160~$\mu$m maps, and especially with the 70~$\mu$m map. 

The 13.5~{GHz} vs.\ 24~$\mu$m data correlation is, at first glance, more pronounced. However, this behavior is not 
straightforward to interpret. As discussed by \cite{pal12}, the 24~$\mu$m emission arising from the interior of HII regions, 
contrary to what happens in the diffuse ISM, is not necessarily associated with VSG emission, but rather with BG replenishment,
 as proposed by \cite{eve10} for the case of wind-blown bubbles. Therefore, for 24~$\mu$m, the correlation in 
Figure \ref{fig:ttplot13vs160_70_24} is not a priori an indication that the AME carriers are small dust grains.

\begin{table}
\caption{Summary of the correlation analysis of the 13.5~{GHz} map vs.\ the {IR} maps. We
 provide: the Pearson correlation coefficients, $\rho$; PTE, {i.e.,} the correlation value in the 
case of null correlation; at each wavelength, the slope of the linear fit to the 13.5~{GHz} map vs.\  
{IR} map distribution. The slope is reported only when PTE~$< 0.002$.} 
\label{tab:pearson2}
\footnotesize
\begin{tabular}{l l c c l }
IR map & Region & $\rho$ & PTE & slope  \\
\hline\hline

\Spitzer 24 & G29.0-0.6  &   0.70285 &   0.00000 &    1.19 $\pm$    0.18)~10$^{-3}$ \\
\Spitzer 24 & G29.1-0.6  &   0.42984 &   0.00354 & ....\\ 
\Spitzer 24 & G29.1-0.7  &   0.13497 &   0.14381 &  ....\\ 

\hline

PACS 70 & G29.0-0.6  &   0.41135 &   0.00184 &    (8.9 $\pm$    2.9)~10$^{-5}$ \\
PACS 70 & G29.1-0.6  &   0.63511 &   0.00001 &    (8.9 $\pm$    7.6)~10$^{-4}$ \\
PACS 70 & G29.1-0.7  &  -0.05484 &   0.33345 &  ....\\ 

\hline

PACS 160 & G29.0-0.6  &   0.21876 &   0.06761 &  ....\\ 
PACS 160 & G29.1-0.6  &   0.60239 &   0.00003 &    (3.98 $\pm$    0.88)~10$^{-4}$ \\
PACS 160 & G29.1-0.7  &  -0.17157 &   0.08762 &  ....\\ 

\hline

SPIRE 250 & G29.0-0.6  &   0.02612 &   0.43006 &  ....\\ 
SPIRE 250 & G29.1-0.6  &   0.48720 &   0.00096 &    (4.8 $\pm$ 1.4)~10$^{-4}$ \\
SPIRE 250 & G29.1-0.7  &  -0.22104 &   0.03961 &  ....\\ 

\hline

SPIRE 350 & G29.0-0.6  &  -0.03789 &   0.39911 &  ....\\ 
SPIRE 350 & G29.1-0.6  &   0.42209 &   0.00415 &  ....\\ 
SPIRE 350 & G29.1-0.7  &  -0.21231 &   0.04607 &  ....\\ 

\hline

SPIRE 500 & G29.0-0.6  &  -0.03584 &   0.40445 & ....\\ 
SPIRE 500 & G29.1-0.6  &   0.41731 &   0.00457 &  ....\\ 
SPIRE 500 & G29.1-0.7  &  -0.19344 &   0.06282 &  ....\\ 

\hline

\hline

\end{tabular}
\end{table}

In addition to the IR maps discussed above, we incorporated the dust parameter maps derived by \cite{tib12}, 
using the  \dustem~dust model \citep{com11}, in the correlation analysis. These parameter maps include important physical quantities such as: (i) 
$Y_{\rm VSG}$, the abundance of VSGs relative to BGs; (ii) $Y_{\rm PAH}$, the abundance of PAHs relative
 to BGs; (iii) $\chi_{\rm ERF}$, the strength of the exciting radiation field, as parameterized in 
\cite{tib12}; (iv) $N_{\rm H}$, the hydrogen column density; and (v) $T_{\rm EQ}$, the dust equilibrium 
temperature derived as the median temperature of the BGs.

\begin{figure}
\epsscale{1.0}
\plotone{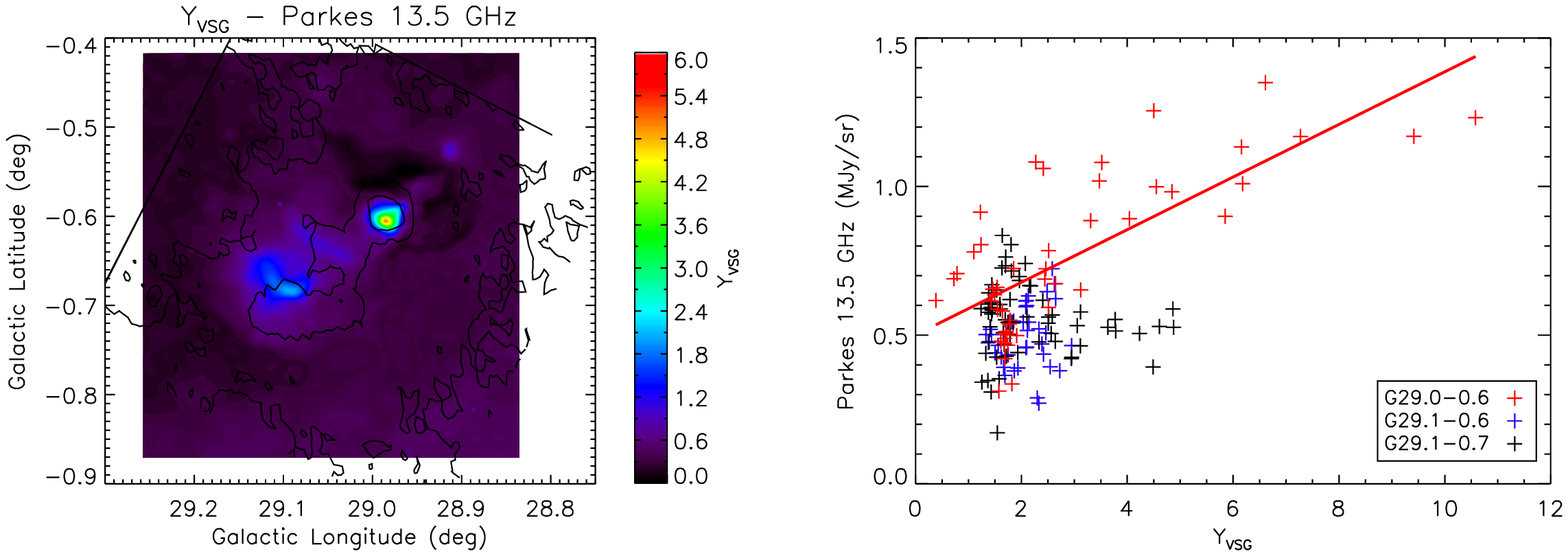}
\plotone{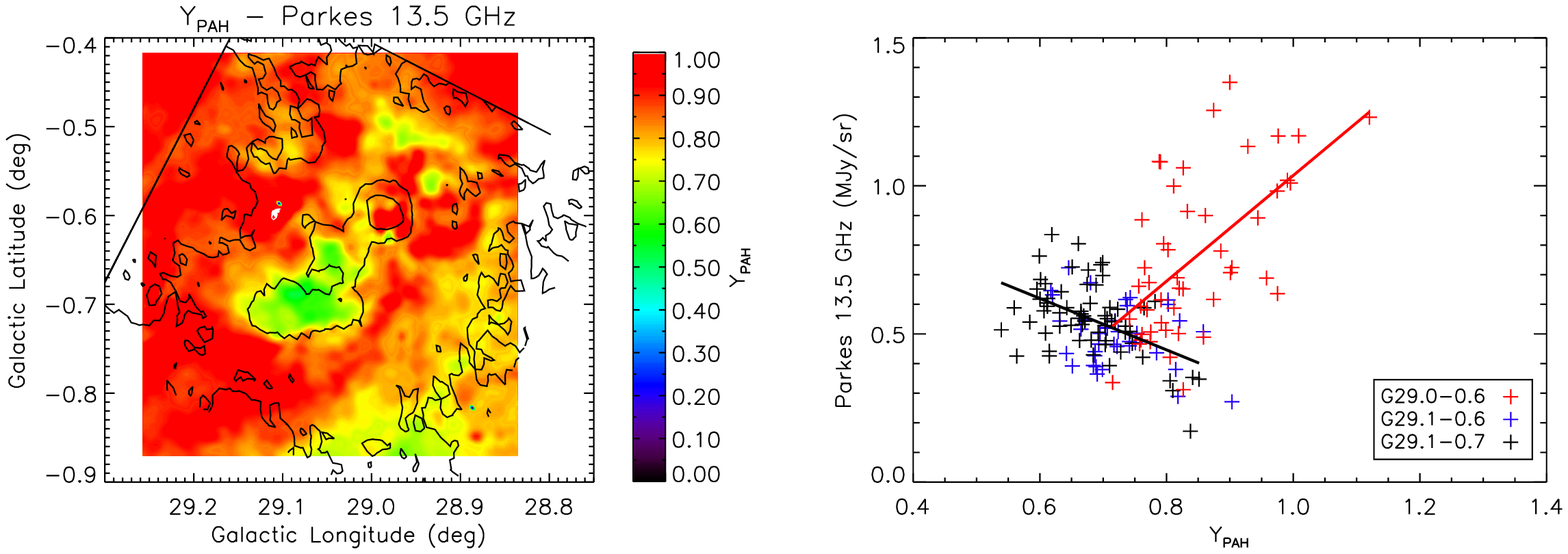}
\plotone{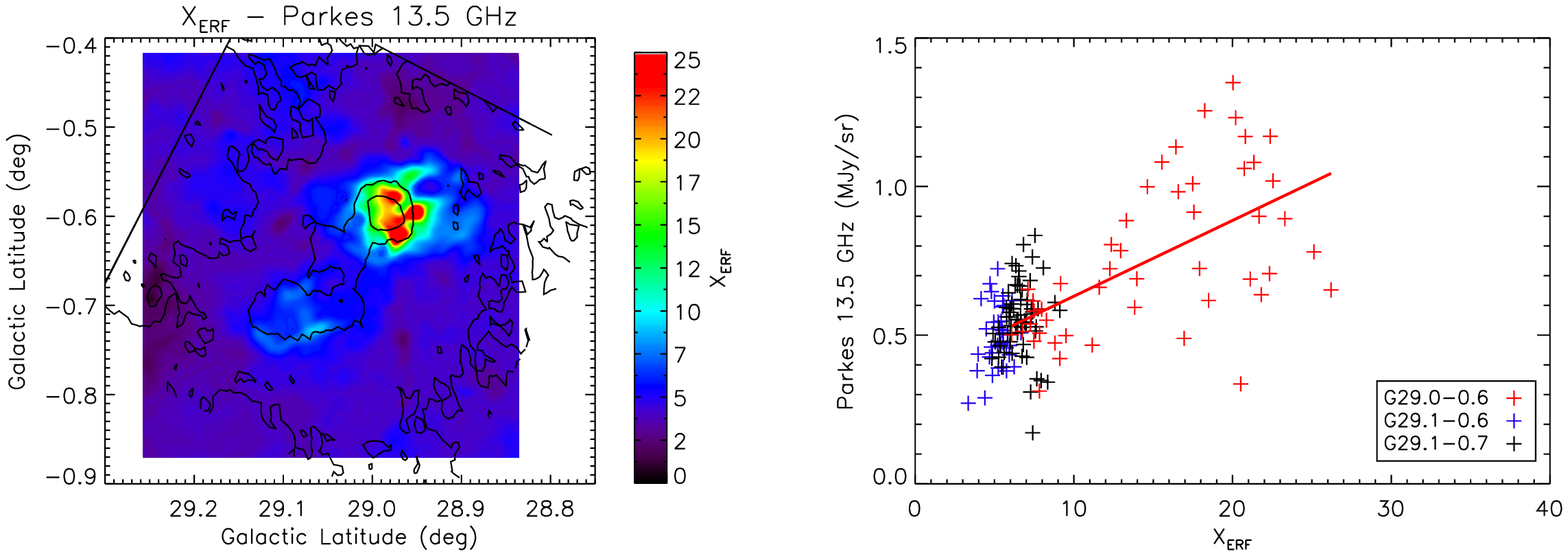}
\caption{Ku-band 13.5~{GHz} map vs.\ \dustem~$Y_{\rm VSG}$, $Y_{\rm PAH}$ and $\chi_{\rm ERF}$ maps (top to bottom).
The \dustem~parameter maps, with the 13.5~{GHz} in contours (with levels at 0\%, 33\% and 66\% of the peak values), are shown on the left. The corresponding 
scatter plots, {i.e.,} {microwave} map vs.\ parameter map, are shown on the right. Different colors
 identify different regions, following the color scheme in Figure~\ref{regions}.} 
\label{fig:ttplot13vsvsg_pah_erf}
\end{figure}

\begin{figure}
\epsscale{1.0}
\plotone{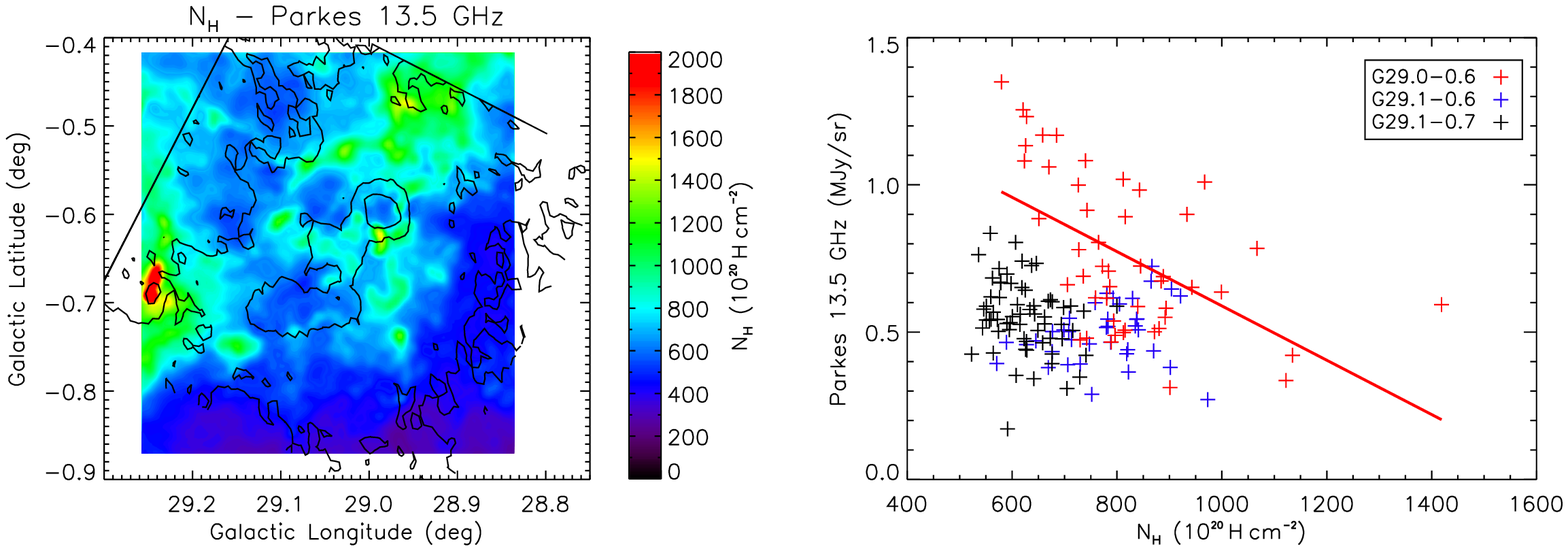}
\plotone{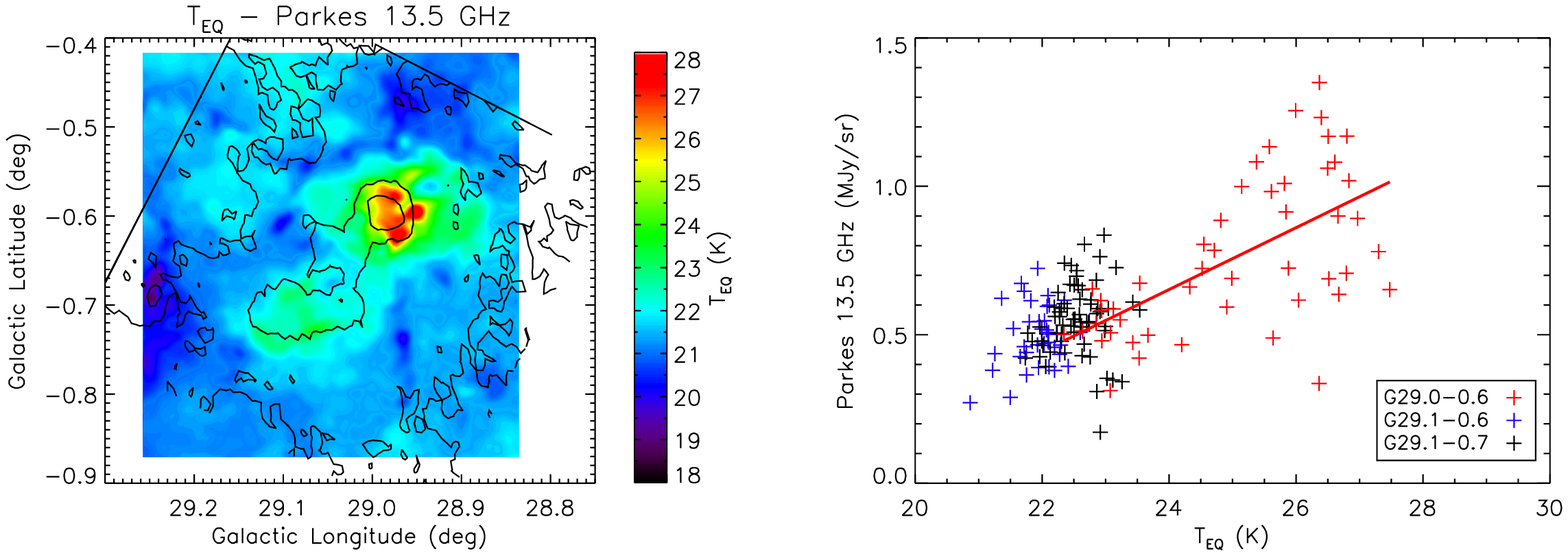}
\caption{Ku-band 13.5~{GHz} map vs.\ \dustem~$N_{\rm H}$ and $T_{\rm EQ}$ maps (top to bottom). The \dustem~parameter
 maps, with  the 13.5~{GHz} in contours (with levels at 0\%, 33\% and 66\% of the peak values), are shown on the left. The corresponding scatter plots, 
{i.e.,} {microwave} map vs.\ parameter map, are shown on the right. Different colors identify
 different regions, following the color scheme in Figure~\ref{regions}.} 
\label{fig:ttplot13vsnh_teq}
\end{figure}

In Figure~\ref{fig:ttplot13vsvsg_pah_erf}, and \ref{fig:ttplot13vsnh_teq}, we compare the RCW175
 13.5~{GHz} map with the parameter maps derived from \dustem: on the left, we provide the
 parameter maps and, as contours, the microwave map, with all the maps at a common angular 
resolution; on the right, we show the corresponding scatter plot, {i.e.,} 13.5~{GHz} 
data vs.\ parameter map, where, once again, different colors denote different regions of the map. 

We have calculated the Pearson correlation coefficients, $\rho$, for the 13.5~{GHz} map and
 the various IR or \dustem~parameter maps, together with the probability to exceed (PTE) that 
correlation value in the case of null correlation. In doing so, we have assumed that the variable 
$t= \rho \sqrt((n-2)/(1-\rho)^2)$ follows a \textit{t}-Student distribution,  with $n-2$ degrees of 
freedom, $n$ being the number of data points.\footnote{The Student \textit{t} distribution is a very 
good approximation of the underlying distribution of the correlation coefficients, particularly
when the correlation coefficient is zero. In this sense it can be used to test the hypothesis that 
the data are uncorrelated. More information on this topic can be found in Chapter 14.15 of~\cite{tKEN77b}.}
The result of this correlation measure is reported in Table~\ref{tab:pearson2} for the correlations between the microwave vs.\  
IR maps, and in Table~\ref{tab:pearson1} for the correlations between the microwave vs.\ \dustem~parameter maps. 
When PTE~$< 0.002$, we also provide the slope of the linear fit to the microwave map vs.\  
parameter map. Microwave vs.\  IR dust emissivities, in terms of microwave brightness temperature relative 
to 100 $\mu$m brightness, are sometimes used to highlight common behavior of AME regions
\cite[see e.g., ][]{dic12}. Comparison with our correlations is not straightforward although we can see 
a similar qualitative behavior. 

For the \dustem~analysis, both the plots and Pearson coefficients reveal a clear correlation between the microwave emission in G29.0$-$0.6 and the dust parameters ($Y_{\rm VSG}$, $Y_{\rm PAH}$, $X_{\rm ERF}$, and $T_{\rm EQ}$). The presence of highly embedded ultracompact HII regions, which could induce the observed correlations in this source component, has been ruled out by the analysis carried out by \cite{tib12}.  In this light, the above correlations indicate that the 13.5~{GHz} emission observed in G29.0$-$0.6 could indeed be ascribed to AME. In the case of the G29.1 system ({i.e.,} G29.1$-$0.6 and G29.1$-$0.7), no significant correlation is found. On the contrary, we find some indication of an anti-correlation between the microwave emission and $Y_{\rm PAH}$.  This result is in apparent contradiction with the one for G29.0$-$0.6. Without additional information, we can only speculate that, in this case, the 13.5~{GHz} emission may have an origin other than dipole emission from PAHs or VSGs. 

Interestingly, our analysis highlights, both for G29.0$-$0.6 and the G29.1 complex, also an anti-correlation between the spatial distribution of  AME and $N_{H}$. Similar behaviour has been observed by \cite{lag03} and \cite{vid11}. We can interpret this result as evidence that the bulk of AME is not generated in the PDR, where the peak of $N_{H}$ is reached along the line of sight, rather in the ionized interior of the HII region (in G29.0$-$0.6) or in the surrounding diffuse medium (i.e. G29.1$-$0.1 and G29.1$-$0.7). In these regions, the abundance of PAHs and VSGs will be lower with respect to the abundance in the PDR, especially in the interior of G29.0$-$0.6, due to radiation pressure drift, as investigated by \cite{dra11}. However, our result suggests that, despite a decreased abundance, mechanisms such as ion collisions and plasma drag may supply the necessary momentum to the grains to allow them to spin, and hence emit efficiently at microwave wavelengths.

\begin{table}
\caption{Summary of the correlation analysis of the 13.5~{GHz} map vs.\ the \dustem~parameter maps. 
We provide: the Pearson correlation coefficients, $\rho$; PTE, {i.e.,} the correlation value in the 
case of null correlation; for each parameter, the slope of the linear fit to the 13.5~{GHz} map vs.\  
\dustem~parameter map correlation. The slope is reported only when PTE~$< 0.002$.} 
\label{tab:pearson1}
\footnotesize
\begin{tabular}{l l c c r l }
Parameter & Region & $\rho$ & PTE & slope & \\
\hline\hline

$Y_{VSG}$ & G29.0-0.6  &   0.74537 &   0.00000 &    (8.8 $\pm$    1.2)~10$^{-2}$  &MJy/sr \\
$Y_{VSG}$ & G29.1-0.6  &   0.17953 &   0.14040 &    ....\\ 
$Y_{VSG}$ & G29.1-0.7  &  -0.10647 &   0.20120 &   ....\\ 

\hline

$Y_{PAH}$ & G29.0-0.6  &   0.59550 &   0.00000 &    (1.78 $\pm$    0.36)  & MJy/sr   \\
$Y_{PAH}$ & G29.1-0.6  &  -0.31103 &   0.02867 &   ....\\ 
$Y_{PAH}$ & G29.1-0.7  &  -0.49451 &   0.00002 &   (-8.7 $\pm$  1.9)~10$^{-1}$  & MJy/sr   \\

\hline

$\chi_{ERF}$ & G29.0-0.6  &   0.57373 &   0.00001 &    (2.55 $\pm$ 0.54)~10$^{-2}$   & MJy/sr \\
$\chi_{ERF}$ & G29.1-0.6  &   0.22391 &   0.08829 &  ....\\ 
$\chi_{ERF}$ & G29.1-0.7  &   0.09043 &   0.23864 &  ....\\ 

\hline

$N_{H}$   & G29.0-0.6  &  -0.53709 &   0.00004 &   (-9.2 $\pm$ 2.1)~10$^{-4}$ & (MJy/sr)/(10$^{20}$H~cm$^{-2}$)  \\
$N_{H}$   & G29.1-0.6  &   0.24156 &   0.07200 &  ....\\ 
$N_{H}$   & G29.1-0.7  &  -0.24629 &   0.02489 &  ....\\ 

\hline

$T_{EQ}$ & G29.0-0.6  &   0.60537 &   0.00000 &    (1.04 $\pm$    0.20)~10$^{-2}$  & MJy/sr  \\
$T_{EQ}$ & G29.1-0.6  &   0.24428 &   0.06970 &  ....\\ 
$T_{EQ}$ & G29.1-0.7  &   0.10209 &   0.21107 &  ....\\ 

\hline

\hline

\end{tabular}
\end{table}

\section{Polarization results}\label{pol}

As described in section \ref{parkes}, we have performed polarization observations of RCW175 at 21.5~{GHz} 
using 1-dimensional scans through the center of the region. Our polarization measurements stand in the low 
signal-to-noise regime (typically $PI/\sigma< 3$). In this case, it is well known \cite[see e.g., ][]{war74,sim85}
 that the determination of reliable polarized fluxes is hindered by the fact that the posterior distribution of the
 polarized intensity, $PI$, does not follow a normal (Gaussian) distribution. Also, $PI$ is a quantity that must
 always be positive, which introduces a bias into any estimate. For any true $PI_0$ we would expect to measure on
 average a polarization $PI>PI_0$. In order to de-bias the measured polarized fluxes, thus obtaining an estimate of
 the underlying real polarization level $PI_0$, we follow the Bayesian approach described in \cite{vai06}, consisting
 of integrating the posterior probability density function over the parameter space of the true polarization. For
 $PI/\sigma<5$ we then integrate the analytical posterior shown in \cite{vai06}, to get the most likely de-biased 
polarized intensity and the associated confidence intervals at the $68.3\%$ level, setting upper limits when 
$PI/\sigma<\sqrt{2}$. When $PI/\sigma\ge 5$ we just obtain the polarized intensity as $\sqrt{PI^2-\sigma^2}$, following
 \cite{war74}. \cite{vai06} focuses on the de-biasing of the polarized intensity, while no analytical solution is
 given for the posterior of the polarization fraction. Then, in order to de-bias the polarization fraction we numerically 
evaluate the corresponding posterior through Monte Carlo simulations, and obtain estimates of $PI/I$ or upper limits 
in an equivalent way as it is done for $PI$. 

\begin{figure}
\epsscale{0.9}
\plotone{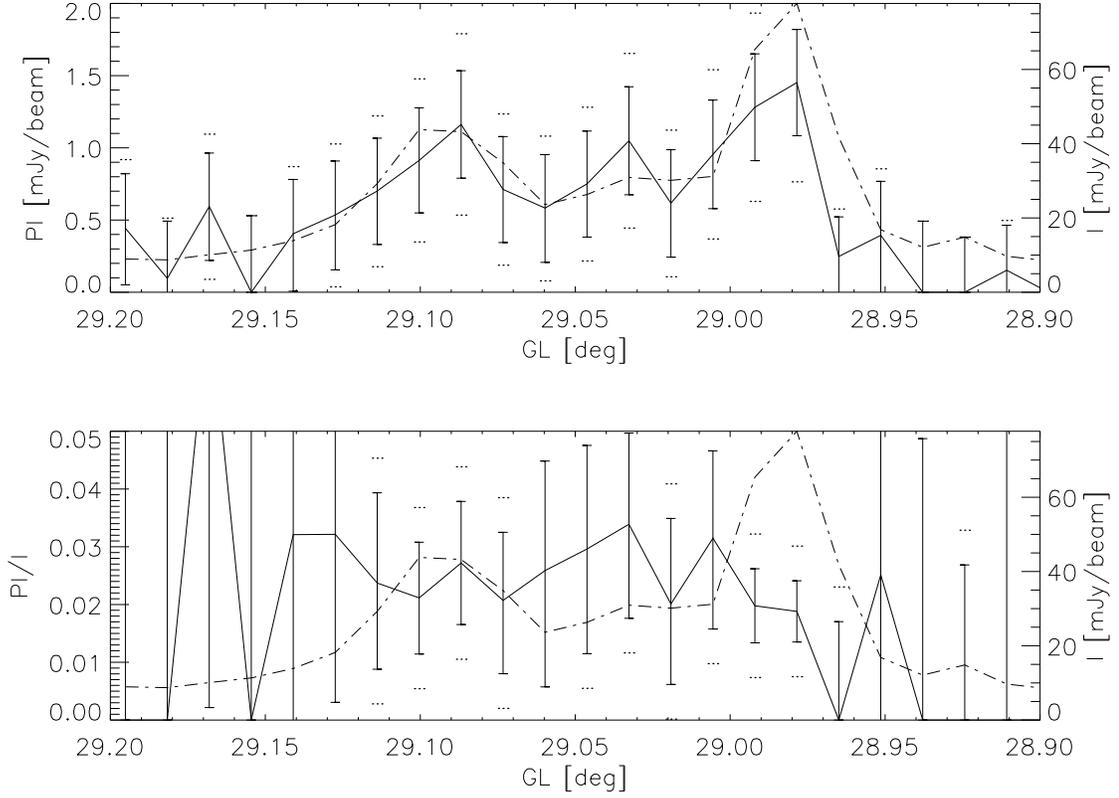}
\caption{Linear polarization measurements performed at 21.5~{GHz}. For polarization measurements, 
1-dimensional scans have been preformed on RCW175 with coordinates shown in Figure \ref{regions}. The top panel displays
the polarized intensity (solid line) along with an indication of the emitted intensity in arbitrary units (dot-dashed line).
The error bars on $PI$ show 2-sigma statistical uncertainties (solid line). The dotted bars show the linearly summed 
statistical uncertainties (2-sigma) to twice the systematic uncertainty (i.e. $2\times11\%$). In the bottom panel we show, with a solid line, the 
relative polarized emission and, with the dot-dashed line we give an indication of the intensity emitted at
 21.5~{GHz}. Also in this plot we include 2-sigma statistical uncertainties arising from $PI$ and $I$ measurements and in the 
dotted bars we linearly added to the 2-sigma statistical uncertainty, twice the residual instrumental polarization 
we qualified with 0.3\%. Measurements are one beam spaced.  \label{pol_plot}}
\end{figure}

The de-biased estimates for the polarized signals are shown in Figure~\ref{pol_plot}. In the top plot we show the 
polarization intensity in units of {mJy/beam}, where the receiver beam at 21.5~{GHz} has a FWHM of 
67~{arcsec}. Overplotted is the intensity measured at the same frequencies. The points in the plot are
 separated by one beam and the errors are quoted at 2-sigma with systematic uncertainty (which includes the calibration, 
the atmospheric opacity and the residual instrumental polarization) linearly added in the dotted error bars. In the lower
 plot we show polarization fraction. Also in this second plot, the dotted error bars include the statistical and the 
systematic uncertainties, in this case being quantified to the level of 0.3\%.

The observed source appears to be weakly polarized to the level of a few per cent, especially in the center of our 
scans. Limiting our analysis to the core of the source ({i.e.,}  for galactic latitude $28\degree.96<GL<29^{\circ}.20$) 
we find a weighted average polarized signal of $2.2\pm0.2 (rand.) \pm0.3 (sys.)$\% (68\% CL), where statistical uncertainties are the result of 
the weighted average within the core of the source and the systematic uncertainty is the residual instrumental polarization 
after correction.

Considering the relatively steep spectrum of the integrated low frequency emission of RCW175, it is 
plausible that some synchrotron contamination is present along the line of sight towards the source accounting for the relatively low level of polarization. We have modeled the presence of synchrotron radiation through the 
low frequency fit preformed in section \ref{sed}. Assuming an average 30\% polarization only for the synchrotron component, 
we expect up to 10\% polarization of the total emission at 21.5~{GHz}. Also, weak polarization could be due to free-free emission in particular specific conditions of optically thick medium towards the edges of the clouds \citep{kea98}. On the other hand, our result is compatible 
with the low level of polarization expected from electric dipole emission (i.e. $\lesssim 1\%$ at the frequencies of our observations; \cite{hoa13}). Multi-frequency polarization measurements would be required 
to discriminate whether the polarized emission comes from the synchrotron or anomalous microwave emission component.

\section{Conclusions}\label{con}

We have observed the HII region RCW175 with the Parkes telescope at 8.4, 13.5, and 21.5~{GHz}. These observations
confirm and strengthen the evidence for AME spatially correlated with thermal dust emission. In particular, approximately 
half of the RCW175 flux density in the range 15--30~{GHz} appears to be due to AME, as previously reported by 
\cite{dic09}. The SED fit suggests that multiple spinning dust components are present in the region, {i.e.,} one
 associated with diffuse ionized gas, and one associated with dense molecular gas. 

Combining the new Parkes observations with ancillary data, we generated an anomalous emission map at 13.5~{GHz} 
for the RCW175 complex. The map shows that the microwave excess is characterised by both a diffuse component and a
 compact structure which overlaps with the location of G29.0$-$0.6. This corroborates the hypothesis that AME regions 
have a spatial extension not coincident to the bulk of microwave emission in a given region, as noted, based on statistical arguments, 
by \cite{pla14a}. 

The cross-correlation analysis of the microwave map with both the IR maps and the \dustem~parameter maps obtained for 
the whole region by \cite{tib12}  highlights a good degree of correlation mainly for G29.0$-$0.6. In this component, 
the investigation on small angular scales reveals an overall decrease of the microwave--IR correlation,  with the exception 
for the 24~$\mu$m map, for which the correlation is more pronounced with respect to other wavelengths. No strong correlation 
is found for the other HII region components ({i.e.,} G29.1$-$0.6 and G29.1$-$0.7). In G29.0$-$0.6, the correlation with short 
wavelength IR data, with PAH abundance, $Y_{\rm PAH}$, with the radiation field strength, $\chi_{\rm ERF}$, and with dust temperature,
 $T_{\rm EQ}$, might indicate that the AME carriers are small dust grains. However, we emphasize that the origin of the 24~$\mu$m emission arising
from the interior of HII regions is still debated. 
 
The observed anti-correlation of  $N_{\rm H}$ with the microwave map, already observed by \cite{lag03} and \cite{vid11}, 
suggests that the bulk of AME is likely emitted in low-density regions, i.e. in our case the interior of G29.0$-$0.6 and the diffuse medium surrounding the G29.1
complex. Additionally, it is important to keep in mind that, if the intensity of the excitation field increases, the ion fraction equally
increases, leading to a higher degree of excitation for PAHs and VSGs, hence to an effective anti-correlation between microwave emission and $N_{\rm H}$. 
In the G29.1 complex, the prevalent diffuse emission and the mild anti-correlation between $Y_{\rm PAH}$ and the 13.5~{GHz} map seems to suggest a
different origin for the observed microwave excess.  We emphasize that part of the  13.5~{GHz}, microwave excess could be due to a rising of the
spectral indices between 1.4~{GHz} and 5~{GHz}, compatible with the presence of strong stellar winds \citep{sca12,pal14} from massive OB stars. Polarization measurements at 21.5~{GHz} show a low level of average polarized signal toward the center of the source of $2.2\pm0.2 (rand.) \pm0.3 (sys.)$\% (68\% CL) compatible with the low level of polarization of the AME or with a residual astrophysical contamination due to synchrotron or free-free from the observed region. This aspect of the analysis demands further investigations in the future.

\acknowledgments

We acknowledge the logistic support provided by Parkes operators. The Parkes radio telescope is part of the Australia 
Telescope National Facility, which is funded by the Commonwealth of Australia for operation as a National Facility managed 
by CSIRO. We thank the referee for constructive comments that improved the paper.

\clearpage

\end{document}